\documentclass[a4paper,12pt]{article}
\usepackage[english]{babel}
\usepackage{url}
\usepackage{layaureo}
\usepackage[dvips]{epsfig}

\usepackage{graphicx}
\usepackage{color}
\usepackage[nomarkers,nolists]{endfloat}
\usepackage{multirow}

\usepackage{amsthm}
\usepackage{amsopn}
\usepackage{amsmath}
\usepackage{amsfonts}
\usepackage{amssymb}
\usepackage{graphicx}
\usepackage{setspace}
\usepackage{natbib}
\usepackage{booktabs}
\usepackage[colorlinks,breaklinks]{hyperref}
\hypersetup{
    colorlinks,%
    citecolor=black,%
    filecolor=black,%
    linkcolor=black,%
    urlcolor=black
}
\usepackage{lscape}
\usepackage{ctable}

\usepackage{array}
\newcolumntype{C}[1]{>{\centering\let\newline\\\arraybackslash\hspace{0pt}}m{#1}}

\numberwithin{equation}{section}

\newcommand{\R}{\mathbb{R}}

\DeclareMathOperator{\Tr}{tr}

\title{Are multi-factor Gaussian interest rate term structure models still useful? An empirical analysis on Italian BTPs}
\author{
Michele Leonardo Bianchi\\
}

\begin{document}

\begin{center}
{\noindent \textbf{\Large Are multi-factor Gaussian term structure models still useful? An empirical analysis on Italian BTPs}}

\vspace{14pt}


{\noindent Michele Leonardo Bianchi\textsuperscript{a,}\footnote{ This publication should not be reported as representing the views of the Banca d'Italia. The views expressed are those of the author and do not necessarily reflect those of the Banca d'Italia.}}


{\noindent \small\textsuperscript{a}\it Regulation and Macroprudential Analysis Directorate, Banca d'Italia, Rome, Italy}\\
%

\vspace{20pt}

This version: \today

\end{center}

\vspace{14pt}

\noindent {\bf Abstract.} In this paper, we empirically study models for pricing Italian sovereign bonds under a reduced form framework, by assuming different dynamics for the short-rate process. We analyze classical Cox-Ingersoll-Ross and Vasicek multi-factor models, with a focus on optimization algorithms applied in the calibration exercise. The Kalman filter algorithm together with a maximum likelihood estimation method are considered to fit the Italian term-structure over a 12-year horizon, including the global financial crisis and the euro area sovereign debt crisis. Analytic formulas for the gradient vector and the Hessian matrix of the likelihood function are provided.

\vskip 0.6cm


\vspace{15pt}

%
%

\newpage

\section{Introduction}\label{sec:Introduction and literature review}

This paper provides a detailed comparative assessment of a number of affine interest rate models applied to the term structure of Italian government securities.

Having reliable estimates of these models allows to conduct scenario analyses that can be used by banks for risk management purposes, e.g. to perform a risk-return analysis or to compute the value-at-risk of a bond portfolio. Furthermore, as indicated by \cite{abdymomunov2014stress}, instead of assuming deterministic scenarios, supervisory authorities may consider hypothetical stochastic scenarios to stress test bank balance sheets. 

The present paper contributes to the relevant literature from three different perspectives: (i) systematically assessing different specifications of affine models, applied to data from one of the largest and most liquid European government bond markets over a 12-year horizon which includes structural breaks; (ii) providing the reader with a detailed tutorial of the technicalities related to the practical implementation of such models (especially in the Appendix);\footnote{ The Matlab code on which this paper is based is available upon request.} (iii) providing new insight into the modeling of the Italian term structure of interest rates, relying on a well-established and simple mathematical framework.  

Multi-factor models have been widely studied in the literature (see \cite{ait2010estimating} and \cite{duffee2012estimation}) for a complete overview). \cite{duan1999kalman} analyze monthly yield series for the U.S. Treasury debt securities with maturities 3, 6, 12 and 60 months spanning the period form April 1964 to December 1997 and conclude for a rejection of exponential affine models. \cite{geyer1999state} have calibrated and tested multi-factor versions of the Cox-Ingersoll-Ross (CIR) model on U.S. interest rates observed monthly from January 1964 to December 1993. Although the specification of multi-factor CIR models is sufficiently flexible for the shape of the term structure, they find strong evidence against the adequacy of the CIR model. They have shown that in the two-factor model the first factor corresponds to the general level of interest rates and the second to the spread between long and short-rates. Therefore, they conclude that the assumption that all factors follow a square-root process 
hinders the explanatory power of additional factors (as it restricts admissible values for 
these factors to be nonnegative). This assumption also affects the decision about the number of factors to be included. As a consequence, they suggest to relax the square-root assumption for all factors, in particular in the case of multi-factor models. \cite{dejong2000affine} shows that a three-factor affine model is able to provide an adequate fit of the cross-section and the dynamics of the term structure. The three factors can be given the usual interpretation of level, steepness, and curvature. \cite{almeida2005note} explains the relation between principal components obtained assuming no dynamic restrictions and the dynamic factors estimated using multi-factor Gaussian term structure models. The author finds that the linear structure embedded in dynamic affine term structure models directly translates into an approximation of the non-negligible principal components by linear transformations of the state vector. \cite{date2009gaussian} have studied the out-of-sample forecasting ability of linear Gaussian 
interest rate models with unobservable underlying factor by considering both the Kalman filter and a non-linear filter. Recently, \cite{rebonato2014principal} have investigated the theoretical framework for the affine evolution of mean-reverting principal component models, and they have shown how the model can be calibrated using both risk-neutral and historical measure information.

As regards the choice of relying on short rate processes, note that neither the Libor nor the swap market models (see \cite{brigo2006interest} and \cite{dempster2014}) can be applied in our context, for two main reasons. First, these market models have been developed to calibrate the quotes of the most liquid interest rate derivatives traded in the market (i.e., caps, floors and swaptions), with the purpose of pricing and hedging exotic derivatives, usually traded over the counter. Even if the information derived by the implied volatilities of these liquid derivatives may be considered to capture market expectations on future rates, the analysis of these derivatives is beyond the scope of this paper. Second, these models are usually calibrated under a static perspective, in the sense that at each given point in time model prices have to be as close as possible to the prices traded in the market. Even if the market is quoting unreasonable prices, the trader has to find the parameters that replicate those 
prices. As explained by \cite{nawalkha2011irm}, traders need to achieve static consistency in order to provide at each given point in time arbitrage-free prices or to hedge a derivatives portfolio. Conversely, our model is calibrated following a dynamic approach, in order to analyze the behavior of yield curves over time. In this respect, we assume that the structural parameters remain constant over time, i.e. we assume that the models considered are time homogeneous while fitting them to observed bond yield curves, taking into account both the cross sectional and the time-series dimensions of Italian sovereign bond rates (i.e. pooling yield curves in a single panel dataset). 

Similar to \cite{bams2003direct} and \cite{ang2013systemic}, we analyze multi-factor Gaussian term structure models under the risk-neutral measure, where the dynamics of the short-rate factors is defined as the sum of Vasicek and CIR factors. Note that the dynamics of the factors of the short rate process under the historical measure, as well as the associated market price of risk, is left unspecified and is not investigated in this paper. That is, only the risk-neutral dynamics is inferred by the observed bond yield curves. Indeed, as the factors driving the short rate process are non-tradable assets, it is not possible to perform a double-calibration in which the historical and risk-neutral model parameters can be jointly estimated. A joint calibration is possible if both the underlying and the derivative are observable, such as for example in the case of equity options (see e.g. \cite{tb2014ijtaf} and \cite{bianchi2017forward} for an example of double-calibration). As a consequence, it is not possible to check whether a given 
historical dynamics of the factors driving the short rate process is feasible or not, and for this reason we prefer not to follow this path. Under the taxonomy introduced by \cite{nawalkha2010tax}, the model we analyze can be defined as {\it single-plus}.



Fitting sovereign bond yields without modeling risk premia nor regime shifts is particularly challenging while analyzing Italian government bond yields over a decade which includes the euro area sovereign debt crisis. In this sense, the comparative analysis of affine models presented in this paper can be viewed as a  ``stress test" of widely used models, aiming to highlight their strengths as well as their limitations. At the same time, the analysis contributes to the literature on Italian government securities with a thorough assessment of different interest models, indicating some viable approaches along with possible pitfalls of well-known models. In the relevant literature, the Italian government bond market has been studied under different perspectives. \cite{barone1989td} tested the CIR model using the prices of Italian government bonds in the secondary market in order to obtain useful indications regarding the efficiency of the secondary market and the consistency between the primary and the secondary 
markets. The authors have conducted a static calibration and have assessed the parameter stability. \cite{decclesia1994immunization} developed a multi-factor model for the yields of Italian bonds and, to control the sensitivity of portfolio returns to movements in the risk factors, implemented appropriate factor immunization models. Based on the observation that short-term bonds interest-rates are characterized by a trend basically driven by the European Central Bank (ECB) key interest rates, \cite{maggi2008assessing} have proposed to filter out the effect of the ECB key interest rates from the term structure of Italian government bonds first, and then, for each maturity, to estimate
a CIR model on the transformed data. \cite{musti2008expectation} have investigated the informational content of the yield curve in the European market using data on the Italian term structures and tested the expectation hypothesis theory. More recently, \cite{impenna2013td} have conducted and analysis on two regulated markets supervised by the Bank of Italy in cooperation with Consob (Commissione Nazionale per la Societ\`a e la Borsa -- Companies and Stock Exchange Commission). They analyzed the price discovery process and the informational role of trading in the interdealer business-to-business (B2B) trading venue (MTS cash), and the business-to-customer (B2C) BondVision market. 

Finally, the present paper contributes to the literature with a view at the implementation of the models analyzed, providing a detailed description of the mathematical details that practitioners have to work out to efficiently implement those models. The dynamic models we study can be casted in a state-space form, where the state is given by the unobservable factors and the observations are given by the term structure. A state and parameter estimation can then be obtained using the Kalman filter together with a maximum likelihood estimation method. However, finding an optimal solution to this type of problems is non trivial and optimization packages should be handled with care. In particular, in order to have reliable optimization algorithms and to compute the standard errors of the estimates exactly, we compute the analytic expression for the gradient vector and the Hessian matrix of the likelihood function considered in the factors and parameter estimation. Full details are provided in the Appendix. 

The remainder of the paper is organized as follows. Section \ref{sec:Model} reviews the Vasicek and CIR multi-factor models considered in the empirical study, including the derivation of bond prices and yields in these models. The estimation algorithms together with the main empirical results are discussed in Section \ref{sec:Fitting}. Section \ref{sec:Conclusions} concludes. The Appendix \ref{sec:Appendix} provides analytic formulas for the gradient and the Hessian of the likelihood function. 

\section{The model}\label{sec:Model}

\subsection{Evaluate zero-coupon bonds}\label{sec:DefaultProbabilities}

We consider the {\it short-rate} approach to model the dynamics of the Italian interest rate term structure. There is a general consensus in assuming a stochastic short-rate instead of a deterministic short-rate to model uncertainty about the future dynamics of the interest rate and credit risk of a given bond. The building block of short-rate models is the integrated process $R_t$, i.e.
$$
R_t = \int_0^t r_s\,ds
$$
where $r_t$ is the instantaneous short-rate process, that is generally assumed to be a stationary affine process. It follows that the price at time $0$ of a zero coupon bond maturing at $t$
\begin{equation}\label{eq:ZCBprice}
B(0,t) = \mathbf{E}\left[\exp({-R_t})\right] = \phi_{R_t}(i),   
\end{equation}
where the expectation is taken under the risk-neutral measure, $\phi_{R_t}(u)$, with $u\in\mathbb{C}$, is the characteristic function of the random variable $R_t$ and $i$ is the imaginary unit. The previous equality easily follows from the definition of the characteristic function of a random variable $X$
$$
\phi_{X}(u) = \mathbf{E}\left[\exp({iuX})\right],
$$ 
where $u\in \mathbb{C}$. Knowing the characteristic function of the process $R_t$, it is straightforward to compute the expectation in equation (\ref{eq:ZCBprice}), and hence the yield to maturity at time 0 of a zero-coupon bond maturing at $t$
\begin{equation*}
y(0,t) = -\frac{1}{t}\log B(0,t) = -\frac{1}{t}\log\left(\mathbf{E}\left[\exp({-R_t})\right])\right) =  -\frac{1}{t}\log\left(\phi_{R_t}(i)\right).   
\end{equation*}
As a consequence, we can compute the entire yield curve by assuming a model for the short rate process $r_t$. In the following, we explore several options, assuming that $r_t$ is a linear combination of stochastic processes that follow either the Vasicek or the CIR model.

\subsection{The Vasicek process}\label{sec:Vasicek}

As proposed by \cite{vasicek1977equilibrium}, a Gaussian Ornstein-Uhlenbeck (OU) process can be used to model the dynamics of the instantaneous short-rate, that is
\begin{equation}
dr_t = \kappa(\eta - r_t)dt +\vartheta dW_t, 
\end{equation}
with $r_0>0$, $\kappa$ and $\vartheta$ positive parameters, $\eta\in\R$, and $W_t$ is a Brownian motion. Usually, the parameter $\eta$ is chosen to be positive. In order to reduce the number of parameters, in our empirical study we will assume that some Vasicek factors will have $\eta = 0$. We recall that, even if $\eta$ is positive, the process can be negative with positive probability (see \cite{brigo2006interest}). The conditional density $p(r_{t+1}|r_t)$ is normal with
\begin{gather*}
E[r_{t+1}|r_t] = \eta(1 - e^{-\kappa \Delta t}) + e^{-\kappa \Delta t}r_t\\ 
Var[r_{t+1}|r_t] = \frac{\vartheta^2}{2\kappa} (1 - e^{-2\kappa \Delta t}).
\end{gather*}
Under the Vasicek model there exists a closed-form expression for the characteristic function of the integrated process (see Proposition 2.6.2.1 in \cite{jeanblanc2009mathematical}) and the integral in equation (\ref{eq:ZCBprice}) can be computed  as follows
\begin{equation}\label{eq:ZCBpriceVasicek}
B(0,t) =e^{A(t) + B(t)r_0},
\end{equation}
where 
\begin{gather*}
A(t) = -\eta t + \eta\frac{1 - e^{-\kappa t}}{\kappa} - \frac{\vartheta^2}{4\kappa^3}\left(1 - e^{-\kappa t}\right)^2 +  \frac{\vartheta^2}{2\kappa^2}\left(t - \frac{1 - e^{-\kappa t}}{\kappa}\right),\\
B(t) = -\frac{1 - e^{-\kappa t}}{\kappa}.
\end{gather*}
For a description of the possible shape of the yield curve under the Vasicek model see \cite{zeytun2007comparative} and \cite{keller2008yield}. 

\subsection{The Cox-Ingersoll-Ross process}\label{sec:CIR}

A well-known way to model the instantaneous short-rate process is to assume the Cox-Ingersoll-Ross (CIR) dynamics (see \cite{cox1985theory}), by considering the following mean-reverting process 
\begin{equation}
dr_t = \kappa(\eta - r_t)dt +\vartheta\sqrt{r_t}dW_t, 
\end{equation}
under the initial condition $r_0>0$, where $W_t$ is a standard Brownian motion and $\kappa$, $\eta$ and $\vartheta$ are positive parameters 
satisfying the additional condition $2\kappa\eta>\vartheta^2$ in order to ensure that the origin is inaccessible,  i.e. that $r_t$ remains positive for all $t$. The conditional density $p(r_{t+1}|r_t)$ of this process has the following form (see e.g. \cite{brigo2006interest})
$$
p(r_{t+1}|r_t)|_{r_{t+1}=z} = c f_{\chi^2(v,l)} (c z)
$$
where $f_{\chi^2(v,l)}$ is the probability density function of a noncentral $\chi^2$ random variable with parameters $v$ and $l$, and
\begin{gather*}
c = \frac{4\kappa}{\vartheta^2(1- e^{-\kappa\Delta t})}, \quad v =\frac{4\kappa\eta}{\vartheta^2}, 
\quad l = c r_t e^{-\kappa\Delta t}.
\end{gather*}
Strictly speaking, the CIR process is not Gaussian,  since its conditional distribution is not normal. 
However, with a slight abuse of terminology, we will refer to it as a Gaussian process, taking into account that its dynamic is driven by a Brownian motion. It can be proven that the conditional mean and variance are
\begin{gather*}
E[r_{t+1}|r_t] = \eta(1 - e^{-\kappa \Delta t}) + e^{-\kappa \Delta t}r_t\\ 
Var[r_{t+1}|r_t] = \frac{\eta\vartheta^2}{2\kappa} (1 - e^{-\kappa \Delta t})^2 +  r_t\frac{\vartheta^2}{\kappa} (e^{-\kappa \Delta t} - e^{-2\kappa \Delta t}).
\end{gather*}
Under the CIR assumption  there exists a closed-form expression for the characteristic function of the integrated CIR process and the bond price in equation (\ref{eq:ZCBprice}) can be computed  as follows (see \cite{schoutens2009levy}, \cite{teng2013numerical} and references therein for the derivation of the formula) 
\begin{equation}\label{eq:ZCBpriceCIR}
B(0,t) =e^{A(t) + B(t)r_0},
\end{equation}
where 
\begin{gather*}
A(t) = \frac{2\kappa\eta}{\vartheta^2} \left(\log(2) +\log\left(\frac{(b/(\kappa - b))e^{(\kappa + b)t/2}}{a e^{bt} - 1}\right)\right),\\
B(t) = -\frac{2}{\kappa - b} \left(\frac{e^{bt} - 1}{a e^{bt} - 1}\right),\\
a = \frac{\kappa + b}{\kappa - b} \quad\text{and}\quad b = \sqrt{\kappa^2 + 2\vartheta^2}.
\end{gather*}
For a description of the possible shape of the yield curve under the CIR model see \cite{zeytun2007comparative} and \cite{keller2008yield}. 

\subsection{A multi-factor short-rate model}\label{sec:MultiFactorModel}

The extension of equation (\ref{eq:ZCBprice}) to the sum of two or more short-rate factors $r_t^1$, $\dots$, {$r_t^d$} 
is straightforward when pairwise independence is assumed between factors. In this case the formula in equation (\ref{eq:ZCBprice}) becomes
\begin{equation}\label{eq:ZCBpriceMultiFactors}
B(0,t) = \phi_{R_t^1}(i) \cdots \phi_{R_t^d}(i). 
\end{equation}
However, if one considers dependent factors, the decomposition in equation (\ref{eq:ZCBpriceMultiFactors}) no longer holds. Normal-based models may still have a closed-form solution, but in general, if one assumes a richer dependence structure (for example, a multivariate model or a copula allowing for tail dependence), Monte Carlo simulations or numerical methods are needed to evaluate bond prices. \cite{feldhutter2008decomposing} proposed a model with six independent factors to calibrate Treasury bonds, corporate bonds, and swap rates using both cross-sectional and time-series properties of the observed yields. This indepedence assumption may be restrictive, although the advantage is that pricing formulas have explicit solutions, and the model is more parsimonious with fewer parameters to estimate. Following a similar approach, in our empirical analysis on the Italian yield curve we will consider different multi-factor models with independent CIR or Gaussian OU (Vasicek) factors. This assumption allows us 
to find a balance between computational tractability and model flexibility. We assume that the short-rate $r_t$  at time $t$ is defined as
\begin{equation}\label{eq:RiskFree}
r_t = \sum_i r_{t}^i,
\end{equation}
where $r^i_t$ may be a CIR or a Vasicek factor. As already observed, when more than one factor is considered, we assume that the Brownian motions driving these factors are independent. We analyze two 1-factor models (1-CIR and 1-Vasicek), three 2-factor models (2-CIR, 2-Vasicek, and 1-CIR\&1-Vasicek), and three 3-factor models (3-CIR, 3-Vasicek, and 2-CIR\&1-Vasicek).

Also in this case, the price of a (defaultable) sovereign zero-coupon bond can be written as
\begin{equation}\label{eq:ZeroCouponPrice}
B(0,t) = \mathbb{E}[e^{-\int_0^t r_s ds}].
\end{equation}
and the model can be calibrated to 
the observed yield curve, that is,
\begin{equation}\label{eq:yield}
y(0,t) = -\frac{1}{t}\log(B(0,t)).
\end{equation}
By equations (\ref{eq:ZCBpriceCIR}) and (\ref{eq:ZCBpriceVasicek}) it follows that the yield is a linear function of $r_0^1$, $\dots$, $r_0^d$,  that is the actual level of the short-rate. As we will see in the following, this property allows one to estimate the model with a Kalman filter.

\section{Fitting market data in practice}\label{sec:Fitting}

A Buono del Tesoro Poliennale (BTP) is an Italian government bond with semi-annual coupon interest payments, principal repaid on maturity and a minimum denomination of EUR 1,000. The main institutional investor secondary market for these sovereign bonds is managed by MTS S.p.a. and the only Italian regulated secondary market for retail investors dedicated to the trading of Italian government securities is managed by Borsa Italiana (MOT). Note that it has been estimated that trading on the electronic platforms for institutional investors, that is outside the MTS market, in 2013 accounted for about 60 per cent of total trading in Italian government securities (see \cite{bi2013fsr}).

We obtain data on Italian government securities from Bloomberg. The zero coupon yield can be obtained by considering the F905 curve series. We considered Italian sovereign bond data from May, 31 2003 to May, 31 2015. The time period in this study includes the high volatility period after the Lehman Brothers filling for Chapter 11 bankruptcy protection (September 15, 2008) and the euro area sovereign debt crisis, during which, in November 2011, the spread between the 10-year Italian BTP and the German Bund with the same maturity was higher than 500 basis point. 

\subsection{Kalman filter}

The stochastic risk-free component $r_t$ is calibrated by fitting the interest rate term structure to the
term structure of Italian government bonds. That is, at each trading day we consider the yield defined in equation (\ref{eq:yield}). There are two possible methodologies to estimate a short-rate model: (1) one can fit the model to the yields observed daily in the market and check both the model flexibility 
and the parameter stability; or (2) one can extract the unobservable  short-rate process (or processes) by using a {\it filter} as described and empirically tested for instance by \cite{chen2003multi} or 
 \cite{osullivan2008parameter}. While the former is a {\it static} estimation,  the latter is a {\it dynamic} approach which captures the behavior of the yield curve over time, taking into account intertemporal consistency of parameter estimates. In this paper we will follow a dynamic approach. In all the cases we are interested in, the model can be written in the following form
\begin{equation}\label{eq:StateSpaceModel}
\begin{split}
x_t &= f(x_{t-1},\Theta,v_{t-1})\\
z_t &= h(x_{t},\Theta,\varepsilon_{t})\\
\end{split}
\end{equation}
where $t$ is the day counter, $x_t$ is the $d$-dimensional state variable (also referred to as the latent or unobservable factor; $r_t$ in the notation of section \ref{sec:Model}), $v_{t-1}$ is the randomness from the state variable with covariance matrix $Q_t$, and $\Theta$ are the parameters. The state variable follows the dynamics described by $f$. The variable $z_t$ represents the set of $n$ observations (in our case the Italian bond term structures observed in the market; $y_t$ in the notation of Section \ref{sec:Model}), while $h$ is the so-called measurement function, which in our case is given by the yield pricing formula \eqref{eq:yield}, and it depends on the state variable, model parameters and measurement noise $\varepsilon_t$.  It is standard to assume that this measurement noise is normally distributed: since we consider more than one yield observations each day, we have a multivariate normally distributed error. Although in general the measurement error covariance matrix $R$ can be a non-
diagonal matrix, for the sake of simplicity it is assumed to be diagonal in this study. Therefore the covariance structure of short-rates is represented only by the model itself and not by the measurement error covariance matrix. Note that, if some component of $x_t$ follows a CIR dynamics, the model proposed in equation (\ref{eq:StateSpaceModel}) is non-Gaussian with respect to the state variable. Nonetheless, since it is linear with respect to the measurement function, the classical Kalman filter can be used (see \cite{chen2003multi}). More details on the maximum likelihood estimation (MLE) method implemented to calibrate market data can be found in \cite{osullivan2008parameter}. Using the notation of \cite{osullivan2008parameter}, the model proposed in equation (\ref{eq:StateSpaceModel}) can be written as
\begin{equation}\label{eq:StateSpaceModelKf}
\begin{split}
x_t &= \Phi^0 e + \Phi^1 x_{t-1} + v_{t-1}\\
z_t &= H^0 e + H^1 x_t + \varepsilon_{t}\\
\end{split}
\end{equation}
where $e$ is a $d$-dimensional column vector with all components equal to $1$, $H^0$ and $H^1$ are $n\times d$ rectangular matrices, $\Phi^0$ and $\Phi^1$  are $d$-dimensional diagonal matrices, and for each maturity $T_j$, with $j=1,\dots,n$, and for each factor $i$, with $i=1,\dots,d$,
\begin{gather}\label{eq:StateSpaceModelKf_input_a}
\Phi^0_{i,i} = (1 - e^{-\kappa_i \Delta t})\eta, \quad \Phi^1_{i,i}  = e^{-\kappa_i \Delta t},
\end{gather}
\begin{gather}\label{eq:StateSpaceModelKf_input_b}
H^0_{j,i} = - \frac{1}{T_j} A_i(T_j), \quad H^1_{j,i}  = - \frac{1}{T_j} B_i(T_j)\quad{\text{and}}\quad \Sigma = \mathrm{diag}(\sigma^2_{\varepsilon})
\end{gather}
where $A_i(t)$ and $B_i(t)$ are defined in equations (\ref{eq:ZCBpriceVasicek}) and (\ref{eq:ZCBpriceCIR}), respectively. Then, we have the diagonal element $i$ of the matrix $Q_t$ equal to
\begin{gather*}
Q_t^i = \frac{\vartheta_i^2}{2\kappa_i} (1 - e^{-2\kappa_i \Delta t}),
\end{gather*}
in the Vasicek case, and
\begin{gather*}
Q_t^i = \frac{\vartheta_i^2\eta_i}{2\kappa_i} (1 - e^{-\kappa_i \Delta t})^2 + r_{t-1}^i \frac{\vartheta_i^2}{\kappa_i}(e^{-\kappa_i \Delta t} - e^{-2\kappa_i \Delta t}),
\end{gather*}
in the CIR case. With a notation similar to \cite{osullivan2008parameter}, the Kalman filter algorithm to calibrate the model defined in equation (\ref{eq:StateSpaceModelKf}) reads as follows. The so-called prediction step is given by
\begin{gather*}
x_t^- = \Phi^0 e + \Phi^1x_{t-1}\\
P_t^- = \Phi^1P_{t-1}(\Phi^1)' + Q_t.
\end{gather*}
The predicted measurement, the prediction error, and its covariance are 
\begin{gather*}
z_t^- = H^0 e + H^1 x_t^-\\
u_t = z_t - z_t^-\\
P_{zz,t} =H^1P_{t}^-(H^1)' + \Sigma.
\end{gather*}
The filtered updates are given by
\begin{gather*}
K_t = P_t^-(H^1)'(P_{zz,t})^{-1}\\
x_t = x_t^- + K_tu_t\\
P_t = (I - K_tH_1)P_t^-.
\end{gather*}
Finally, the log-likelihood function for the maximum likelihood estimation of the model proposed in equation (\ref{eq:StateSpaceModelKf}) is 
\begin{equation}\label{eq:LL}
\log L (\Theta) = -\frac{nd}{2} \log 2\pi - \frac{1}{2}\sum_{t=0}^n \log |P_{zz,t}| - \frac{1}{2}\sum_{t=0}^n u_tP_{zz,t}^{-1}u_t
\end{equation}
where $\Theta$ is the set of model parameters. 

As observed by \cite{chen2003multi}, if one considers a CIR factor, equation (\ref{eq:StateSpaceModelKf}) is an approximation of the the true dynamics, because the innovations in the CIR model have a noncentral $\chi^2$ distributions, in contrast to the normal distribution that is assumed for maximum likelihood estimation of the Kalman filter. For this reason, the model (\ref{eq:StateSpaceModel}) should in principle be calibrated taking into account the non-Gaussian nature of the CIR process. For example, the expectation maximization (EM) algorithm proposed by \cite{schon2011system} could be applied. This approach is not pursued in this paper. 

\subsection{The optimization algorithm}

It is well-known that finding an optimal solution that maximizes the likelihood function in equation (\ref{eq:LL}) is not a simple task. Most of the theoretical literature on this subject does not report explicitly the algorithm applied to solve the optimization problem above nor the computational time needed for the calibration. Note that in optimization packages first and second derivatives are usually approximated with finite differences, if the analytic gradient vector and the Hessian matrix are not provided. In order to speed up the optimization algorithm and to compute the standard error of the estimates in an exact way, we use the analytic expression for the gradient vector and the Hessian matrix (see Appendix \ref{sec:Appendix}){.\footnote{ In a maximum likelihood estimation standard errors can be computed by taking the square root of the diagonal elements of the inverse Hessian matrix at the optimal point.}} As shown in the appendix, the analytical calculation of derivatives involving the CIR factor 
are more cumbersome compared to those involving the Vasicek factor.

In the optimization algorithm we constrain the parameters ($\kappa$, $\eta$, $\vartheta$) of each factor in the region between (1e-4, 1e-4, 1e-4) and (5, 0.1, 0.5) for the CIR factors, or (5, 0.1, 0.1) for the Vasicek factors. The parameter  $\sigma_{\varepsilon}$ ranges between 1e-4  and 0.5.
The optimization algorithm applied in this study is the sequential quadratic programming method implemented in the {\it fmincon} Matlab function in which the option {\it sqp} is selected. The analytic gradient vector of the likelihood is provided in order to speed up the algorithm. As an alternative the interior point algorithm that considers the analytic Hessian matrix can be used. As 
the computation of the analytic Hessian is time-consuming, we only use it to compute the standard errors of the parameters and not in the optimization algorithm. From a practical standpoint, multi-factor models based only on Vasicek factors are simpler to implement, particularly when one wants to use the analytic Hessian matrix. 

As a starting point of the algorithm we select a random point in the parameter region. This approach seems to work for all models except for those involving more than two CIR factors. Given the presence of low (near-zero) rates, the model does not seem to perform well. One should add a negative constant to the short-rate process in equation (\ref{eq:RiskFree}) in order to raise the mean of the factors and give greater flexibility to the model. Alternatively, one can add a Vasicek factor. In this paper, we follow the latter approach. 

\subsection{Empirical results}

In Tables \ref{tab:CalibrationError} and \ref{tab:EstimatedParameters} we report, for each maturity,  parameter estimates and calibration errors for different models of the Italian government bond term structure over the time horizon considered (May 31, 2003 to May 31, 2014). We present results for 
eight different multi-factor short-rate models with CIR and (or) Vasicek factors. In addition, Table \ref{tab:EstimatedParameters} reports the value of the log-likelihood function and its gradient, as well as the first order optimality condition at the optimal point.\footnote{ For the definition of the first order optimality condition we refer to the Matlab documentation.} 

As expected, the analysis shows that the models with more than one factor have a better performance in terms of calibration error. On the one hand, those including a Vasicek factor have a smaller calibration error, due to a higher degree of flexibility. However, the use of the Vasicek factor does not ensure that the short-rate remains positive, even if by construction its mean value is strictly positive. On the other hand, while multi-factor CIR models generate only positive factors, they are less flexible, particularly when interest rates approach the zero lower bound. In this case it may happen that for some multi-factor models the optimal parameters hit the boundaries and this results in a value of the gradient that is far from zero. In addition, for CIR factors some numerical errors may appear in the evaluation of the Hessian and this results in negative values (although small in absolute value) for the square of the standard errors.

To assess the performance of different models, we apply the Akaike information criterion (AIC), defined as
$$
AIC = 2np - 2\log L
$$
where $np$ is the number of parameters and $\log L$ is the model log-likelihood. According to the AIC reported in Table \ref{tab:EstimatedParameters}, the 3-factor Vasicek model performs better than the other models considered in our analysis, i.e. it has the smallest AIC value. The Bayesian information criterion (BIC) gives the same ranking of models is therefore omitted.

\begin{figure}
\begin{center}
\caption[ {\it Estimated factors}]{\label{fig:EstimatedFactors}\footnotesize Estimated factors and short-rate process $r_t$ on the Italy bond yields from May, 31 2003 to May, 31 2014. Recall that the black line is the sum of the unobservable factors, that is the estimated short rate $r_t$. We report the estimated factors for all models investigated. The Kalman filter is considered to extract the unobservable short-rate process.}
\includegraphics[width=\columnwidth]{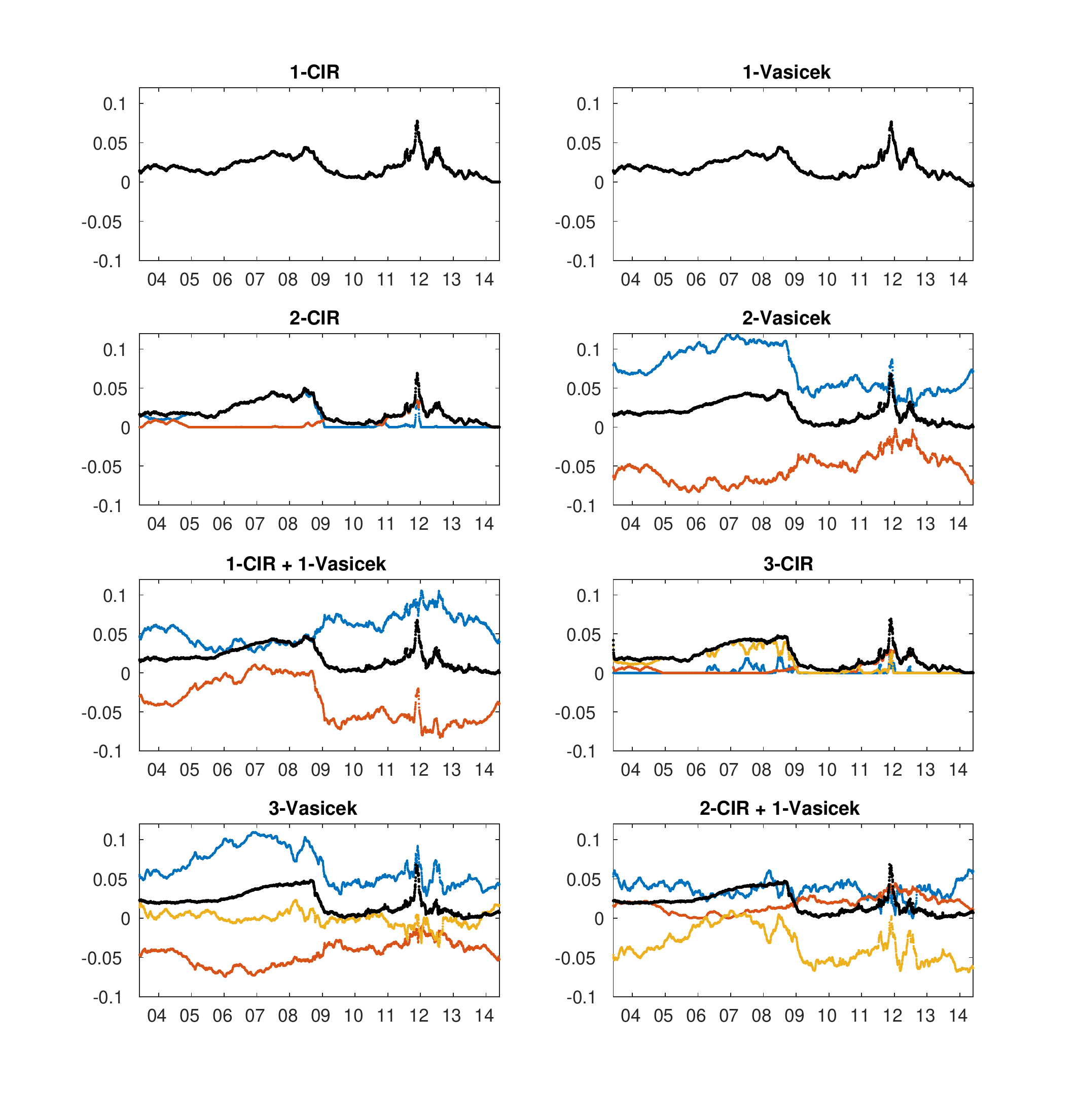}
\end{center}
\end{figure}

Figure \ref{fig:EstimatedFactors} shows the dynamics of the factors for the eight models analyzed in this paper. The value of sum of the factors (in black in Figure \ref{fig:EstimatedFactors}) captures the level of the shortest maturity rates while the other parameters capture the average slope and curvature of the term structure throughout the sample period. It appears clear that if one considers more than one CIR factor, there are factors that hit the zero lower bound. This makes the calibration of these models computationally more challenging, as confirmed by the fact that the optimal parameters depend on the starting point in the optimization. Conversely, if one considers models with at least one Vasicek factor, this becomes negative, even if the sum of the factors (the black line) remains positive. In these cases the optimization algorithm is more robust and the optimal solution does not depend on the initial point.

For comparative purposes, the models performance across maturities and different observation dates is evaluated with the {\it average percentage error} (APE) 
\begin{equation}\label{eq:TotalAPE}
APE = \frac{1}{\bar{y}^{market}}\sum_{t}\sum_{T_i}\frac{|y^{market}_{t,T_i} - y^{model}_{t,T_i}(\theta)
|}{\text{number of observations}},
\end{equation} 
and the {\it root mean square error} (RMSE) 
\begin{equation}\label{eq:TotalRMSE}
RMSE = \sqrt{\sum_{t}\sum_{T_i}\frac{(y^{market}_{t,T_i} - y^{model}_{t,T_i}(\theta))^2}{\text{number of observations}}},
\end{equation} 
where $y_{t,T_i}$ denotes the yield to maturity $T_i$ observed at time $t$, and $\bar{y}^{market}$  is the average yield across time and maturities.

Based on the APE and the RMSE evaluated over the entire sample on successive cross-sections of bond yields multi-factor models perform better than one-factor models. The APE and the RMSE are larger for both CIR and Vasicek 1-factor models and smaller for three-factor models. While the overall APE ranges between 9.88 (1-CIR) and 2.61 per cent (3-Vasicek), the overall RMSE ranges between 48.98 (1-Vasicek) and 18.79 basis points (3-Vasicek). Both APE and RMSE values reported in Table \ref{tab:CalibrationError} show that the calibration error depends on the maturity. For all models, the error is usually larger for the shortest maturities. Beside the CIR model and the Vasicek 1-factor show a similar dynamic for the unobservable factor and a similar calibration error, the estimated parameter differs (see Table \ref{tab:EstimatedParameters}).

\begin{sidewaysfigure}
\begin{center}
\caption[ {\it Italian yield calibration}]{\label{fig:EstimatedYields}\footnotesize Calibration of the best performing short-rate model (3-Vasicek) on the Italy bond yields from May, 31 2003 to May, 31 2014. We report the observed bond rates and the estimated ones for all maturities investigated. The Kalman filter is considered to extract the unobservable short-rate process.}
\includegraphics[width=\columnwidth]{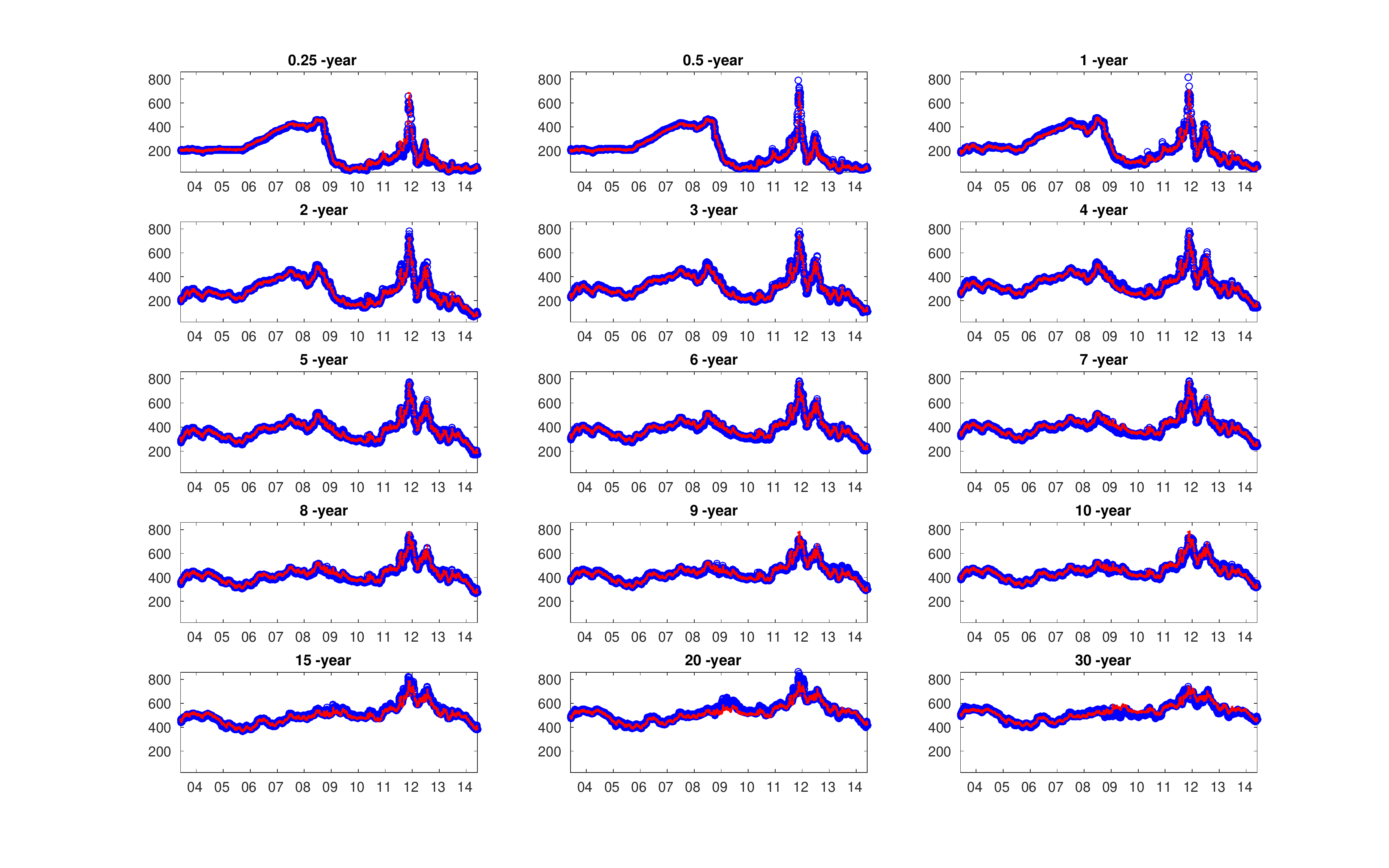}
\end{center}
\end{sidewaysfigure}

\begin{sidewaysfigure}
\begin{center}
\caption[ {\it Italian yield calibration error}]{\label{fig:EstimatedYieldsError}\footnotesize Calibration errors of the best performing short-rate model (3-Vasicek) on the Italy bond yields from May, 31 2003 to May, 31 2014. We report the calibration errors for all maturities investigated. The Kalman filter is considered to extract the unobservable short-rate process.}
\includegraphics[width=\columnwidth]{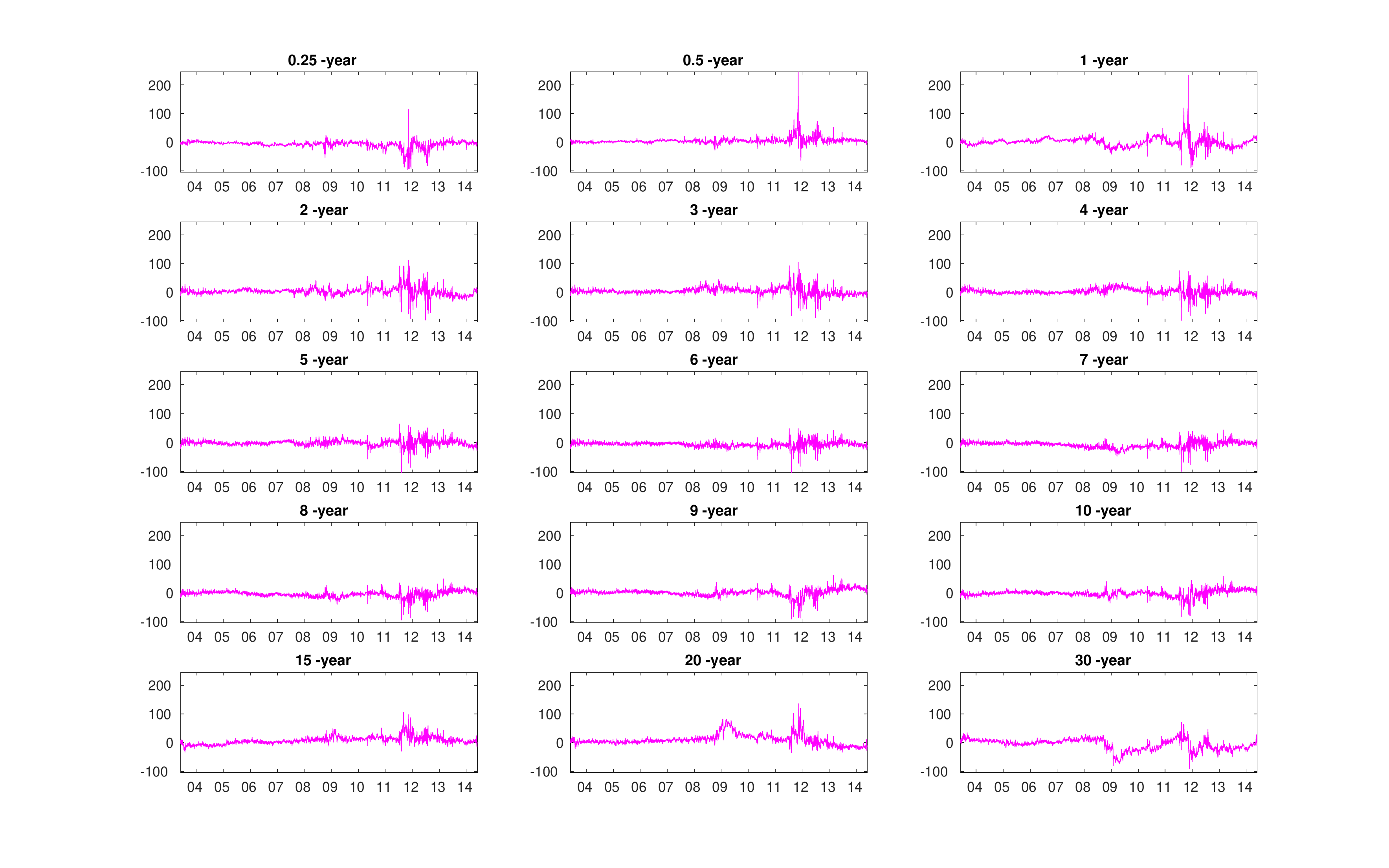}
\end{center}
\end{sidewaysfigure}

In order to have a visual assessment of the fitting exercise, in Figure \ref{fig:EstimatedYields} the estimated rates of the best performing model (i.e. the model with three Vasicek factors) are compared with the market ones. Furthermore, in Figure \ref{fig:EstimatedYieldsError} we show the corresponding calibration errors. On the last trimester of 2011 the model reaches the largest calibration error. We are aware of the fact that Gaussian models are not able to explain sharp movements in the market, even if the performance over the observation period is satisfactory. 

\begin{sidewaysfigure}
\begin{center}
\caption[ {\it Italian yield simulation}]{\label{fig:SimulatedYields}\footnotesize Simulated and observed Italy bond yields. We report the 5 (10, 20, 60, 120 and 260) trading days ahead forecast in the period from from May, 31 2014 to May, 31 2015. For each model, we consider the mean, the 5th and the 95th percentile based on 10,000 simulated scenarios.}
\includegraphics[width=\columnwidth]{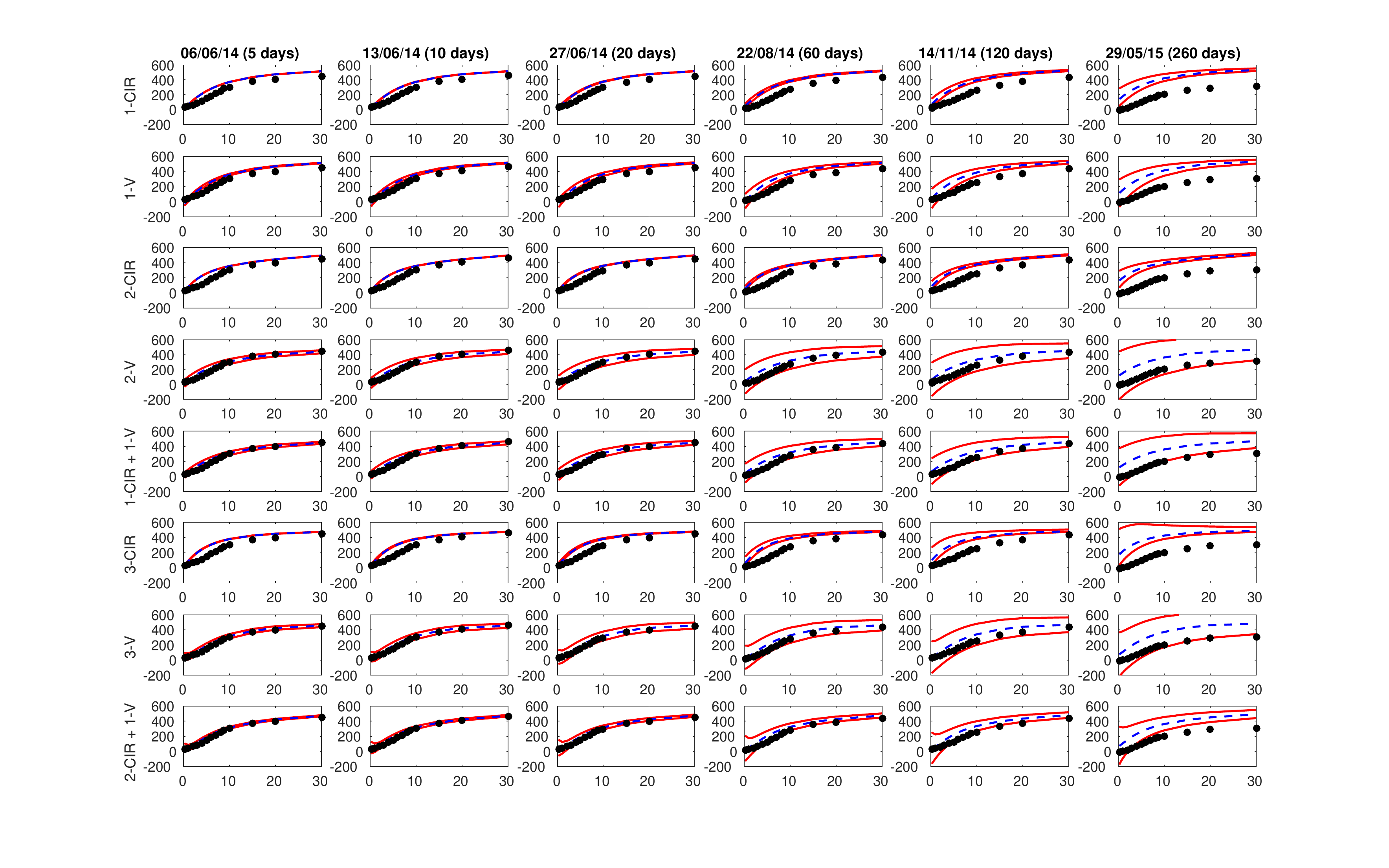}
\end{center}
\end{sidewaysfigure}

In addition to estimating the eight models, we empirically study their forecasting performance over time, by considering 10,000 Monte Carlo simulations. Based on the estimates in the period from May, 31 2003 to May, 31 2014, we compare the simulated yield with the data observed in the market until May, 31 2015. In Figure \ref{fig:SimulatedYields} we show the 5 (10, 20, 60, 120 and 260) trading days ahead forecasts. For each model, we report the mean, the 5th and the 95th percentile based on 10,000 simulated scenarios. Figure \ref{fig:SimulatedYields} shows that the forecasting performance of the 1-factor model is poor. The same is true for the models with only CIR factors. The addition of at least one Vasicek factor in multi-factor models largely improve the forecasting performance, even if negative values for short-term maturities are allowed. As expected, the forecasting performance decreases and the volatility of the forecast increases after 60 trading days. At least for the data considered in this study, 
some models show a satisfactory 3-month ahead forecasts. The 6- and 12-month ahead forecast are far from the yield curve observed in the market. We are aware of the fact that this empirical study can be affected by the expanded asset purchase program announced by the European Central Bank in January 2015.  

\section{Conclusion}\label{sec:Conclusions}

The objective of this paper is twofold. First, we provide a maximum likelihood optimization algorithm based on the Kalman filter in which both the gradient vector and the Hessian matrix can be computed in closed-form. Second, we explore the calibration performance of multi-factor models driven by independent univariate CIR and Vasicek processes under the risk-neutral measure in fitting the Italian term structure. As already observed in other empirical studies, at least three factors are necessary for a satisfactory representation of the behavior of yield curves. This seems a good compromise between a satisfactory performance in terms of calibration error and parameter parsimony. We show that at least a Vasicek factor is needed to properly calibrate the dynamics of the yield curve at least for the data we consider in this paper. If even the Vasicek factor may become negative, the short-rate process, defined as the sum of the factors, remains positive and its expected mean is strictly positive. The calibration 
of multi-factor CIR models is affected by the presence of low level (near zero) rates, and these observed patterns complicate the convergence properties of the optimization algorithm. Finally, the multi-factor models with at least a Vasicek factor seem to show a better forecasting performance.

\newpage

\appendix
  \renewcommand{\theequation}{A.\arabic{equation}}
  \setcounter{equation}{0}  

\section{Appendix}\label{sec:Appendix}

\subsection{Gradient of the likelihood function}\label{sec:GradientKF}

Here we provide the formulas to compute the gradient of the likelihood function to use in the optimization algorithm. For a generic parameter $\xi\in\Theta$, we have that
\begin{equation}\label{eq:gradientLikelihoodKF}
\frac{\partial\log L (\Theta)}{\partial\xi} = -\frac{1}{2}\sum_{t=0}^n\left\{ 2\frac{\partial u_t}{\partial\xi}P_{zz,t}^{-1}u'_t + \Tr\left(P_{zz,t}^{-1}\frac{\partial P_{zz,t}}{\partial\xi}\right) - u_tP_{zz,t}^{-1}\frac{\partial P_{zz,t}}{\partial\xi}P_{zz,t}^{-1}u'_t\right\}.
\end{equation}
The first gradient in equation (\ref{eq:gradientLikelihoodKF}) is given by
\begin{equation*}
\begin{split}
\frac{\partial u_t}{\partial\xi} &\;= - \frac{\partial H^0}{\partial\xi} e -  \frac{\partial H^1}{\partial\xi} x_t^- - H^1\frac{\partial x_t^-}{\partial\xi}\\
\frac{\partial x_t^-}{\partial\xi} &\;= \frac{\partial\Phi^0}{\partial\xi} e + \frac{\partial\Phi^1}{\partial\xi}x_{t-1} + \Phi^1\frac{\partial x_{t-1}}{\partial\xi}.\\
\end{split}
\end{equation*}
In both the Vasicek and the CIR case we have
\begin{equation*}
\frac{\partial\Phi^0}{\partial\xi} =   
\begin{pmatrix}
 \eta\Delta t e^{-\kappa \Delta t} & 1 - e^{-\kappa \Delta t} & 0 \\
\end{pmatrix}'
\end{equation*}
and
\begin{equation*}
\frac{\partial\Phi^1}{\partial\xi} =   
\begin{pmatrix}
-\Delta t e^{-\kappa \Delta t} & 0 & 0 \\
\end{pmatrix}'.
\end{equation*}
The partial derivatives of $H^0$ and $H^1$ are given in Section \ref{sec:I3}. More challenging is the computation of the partial derivatives of $P_{zz,t}$. These involves both the factors parameters and the measurement error parameters.
\begin{equation*}
\frac{\partial P_{zz,t}}{\partial\xi} =  \frac{\partial H^1}{\partial\xi}P_{t}^-(H^1)' + H^1\frac{\partial P_{t}^-}{\partial\xi}(H^1)' + H^1P_{t}^-(\frac{\partial H^1}{\partial\xi})' + \frac{\partial\Sigma}{\partial\xi},
\end{equation*}
where the partial derivatives of $H^1$ are given in Section \ref{sec:I3}, 
\begin{equation*}
\frac{\partial P_{t}^-}{\partial\xi} =  \frac{\partial\Phi^1}{\partial\xi}P_{t-1}(\Phi^1)' + \Phi^1\frac{\partial P_{t-1}}{\partial\xi}(\Phi^1)' + \Phi^1P_{t-1}(\frac{\partial\Phi^1}{\partial\xi})' + \frac{\partial Q_t}{\partial\xi},
\end{equation*}
and the derivatives of $\Sigma$ can be computed by considering that the matrix is diagonal with all element equal to $\sigma^2_{\varepsilon}$. Furthermore, the filtered updates are given by
\begin{equation*}
\begin{split}
\frac{\partial x_t}{\partial\xi} &\;= \frac{\partial x_t^-}{\partial\xi} + \frac{\partial K_t}{\partial\xi}u_t + K_t\frac{\partial u_t}{\partial\xi}\\
\frac{\partial P_t}{\partial\xi} &\;= \frac{\partial P_{t}^-}{\partial\xi} - \frac{\partial K_t}{\partial\xi}H^1P_{t}^- - K_t\frac{\partial H^1}{\partial\xi}P_{t}^- - K_tH^1\frac{\partial P_{t}^-}{\partial\xi},
\end{split}
\end{equation*}
with
\begin{equation*}
\frac{\partial K_t}{\partial\xi} = \frac{\partial P_{t}^-}{\partial\xi}(H^1)'(P_{zz,t})^{-1} + P_{t}^-\frac{\partial (H^1)'}{\partial\xi}(P_{zz,t})^{-1} -  P_{t}^-(H^1)'(P_{zz,t})^{-1}\frac{\partial P_{zz,t}}{\partial\xi}(P_{zz,t})^{-1}.
\end{equation*}
The partial derivatives of $Q_t$ for a Vasicek factor $r$ is
\begin{equation*}
\frac{\partial Q_t}{\partial\xi}  = \frac{\partial}{\partial\xi}\left\{\frac{\vartheta^2}{2\kappa} (1 - e^{-2\kappa \Delta t})\right\} = \frac{\partial q_v}{\partial\xi},
\end{equation*}
where
\begin{equation*}
\begin{split}
\frac{\partial q_v}{\partial\kappa} &\;= -\frac{\vartheta^2}{2\kappa^2} (1 - e^{-2\kappa \Delta t}) + \frac{\vartheta^2 \Delta t}{\kappa} e^{-2\kappa \Delta t}\\
\frac{\partial q_v}{\partial\eta} &\;= 0\\
\frac{\partial q_v}{\partial\vartheta} &\;= \frac{\vartheta}{\kappa} (1 - e^{-2\kappa \Delta t})\\
\end{split}
\end{equation*}
The partial derivatives of $Q_t$ for a CIR factor $r$ is
\begin{equation*}
\begin{split}
\frac{\partial Q_t}{\partial\xi} &\;= \frac{\partial}{\partial\xi}\left\{\frac{\vartheta^2\eta}{2\kappa} (1 - e^{-\kappa \Delta t})^2 + r_{t-1} \frac{\vartheta^2}{\kappa}(e^{-\kappa \Delta t} - e^{-2\kappa \Delta t})\right\}\\
&\;= \frac{\partial}{\partial\xi}\left\{q_{c1} + r_{t-1}q_{c2}\right\}
= \frac{\partial q_{c1}}{\partial\xi} + \frac{\partial r_{t-1}}{\partial\xi}q_{c2} + r_{t-1} \frac{\partial q_{c2}}{\partial\xi}\\
\end{split}
\end{equation*}
where
\begin{equation*}
\begin{split}
\frac{\partial q_{c1}}{\partial\kappa} &\;=  -\frac{\vartheta^2\eta}{2\kappa^2} (1 - e^{-\kappa \Delta t})^2 + \frac{\vartheta^2\eta\Delta t}{\kappa} (e^{-\kappa \Delta t} - e^{-2\kappa \Delta t})\\
\frac{\partial q_{c1}}{\partial\eta} &\;= \frac{\vartheta^2}{2\kappa} (1 - e^{-\kappa \Delta t})^2\\
\frac{\partial q_{c1}}{\partial\vartheta} &\;= \frac{\vartheta\eta}{\kappa} (1 - e^{-\kappa \Delta t})^2\\
\end{split}
\end{equation*}
and
\begin{equation*}
\begin{split}
\frac{\partial q_{c2}}{\partial\kappa} &\;=-\frac{\vartheta^2}{\kappa^2}(e^{-\kappa \Delta t} - e^{-2\kappa \Delta t}) - \frac{\vartheta^2}{\kappa}(\Delta t e^{-\kappa \Delta t} - 2\Delta t e^{-2\kappa \Delta t})\\
\frac{\partial q_{c2}}{\partial\eta} &\;= 0\\
\frac{\partial q_{c2}}{\partial\vartheta} &\;= \frac{2\vartheta}{\kappa}(e^{-\kappa \Delta t} - e^{-2\kappa \Delta t}).\\
\end{split}
\end{equation*}
By considering that the initial point of the KF algorithm are
$$
x_0 = \eta \quad\text{ and }\quad P_0 = \frac{\vartheta^2}{2\kappa}
$$
in the Vasicek case, and
$$
x_0 = \eta \quad\text{ and }\quad P_0 = \frac{\eta\vartheta^2}{2\kappa}
$$
in the CIR case, the derivatives with respect to a generic parameter $\xi$ are simple to compute, that is
in both the Vasicek and the CIR case we have
\begin{equation*}
\frac{\partial x_0}{\partial\xi} =   
\begin{pmatrix}
0 & 1 & 0 \\
\end{pmatrix}
\end{equation*}
and
\begin{equation*}
\frac{\partial P_0}{\partial\xi} =   
\begin{pmatrix}
 -\frac{\vartheta^2}{2\kappa^2} & 0 & \frac{\vartheta}{\kappa}\\
\end{pmatrix}'
\end{equation*}
in the Vasicek case, and
\begin{equation*}
\frac{\partial P_0}{\partial\xi} =   
\begin{pmatrix}
 -\frac{\vartheta^2\eta}{2\kappa^2} & \frac{\vartheta^2}{2\kappa} & \frac{\vartheta\eta}{\kappa}\\
\end{pmatrix}'
\end{equation*}
in the CIR case. These derivatives are needed as initial point of the algorithm that compute the gradient on the likelihood function to maximize.

\subsection{Hessian of the likelihood function}\label{sec:HessianKF}

Here we provide the formulas to compute the Hessian of the likelihood function to use in the optimization algorithm. For generic parameters $\xi_1,\xi_2\in\Theta$, we have that
\begin{equation}\label{eq:hessianLikelihoodKF}
\begin{split}
\frac{\partial\log L (\Theta)}{\partial\xi_1\partial\xi_2} &\;= -\frac{1}{2}\sum_{t=0}^n\bigg\{ 2\frac{\partial u_t}{\partial\xi_1\partial\xi_2}P_{zz,t}^{-1}u'_t - 2\frac{\partial u_t}{\partial\xi_1}P_{zz,t}^{-1}\frac{\partial P_{zz,t}}{\partial\xi_2}P_{zz,t}^{-1}u'_t + 2\frac{\partial u_t}{\partial\xi_1}P_{zz,t}^{-1} \frac{\partial u_t}{\partial\xi_2}'\\
&\quad\qquad\qquad-\Tr\left(P_{zz,t}^{-1}\frac{\partial P_{zz,t}}{\partial\xi_2}P_{zz,t}^{-1}\frac{\partial P_{zz,t}}{\partial\xi_1}\right) + \Tr\left(P_{zz,t}^{-1}\frac{\partial P_{zz,t}}{\partial\xi_1\partial\xi_2}\right)\\
&\quad\qquad\qquad -2\frac{\partial u_t}{\partial\xi_2}P_{zz,t}^{-1}\frac{\partial P_{zz,t}}{\partial\xi_1}P_{zz,t}^{-1}u'_t
 - u_t\frac{\partial}{\partial\xi_2}\Big(P_{zz,t}^{-1}\frac{\partial P_{zz,t}}{\partial\xi_1}P_{zz,t}^{-1}\Big)u'_t\bigg\},
\end{split}
\end{equation}
where
\begin{equation*}
\begin{split}
\frac{\partial}{\partial\xi_2}\Big(P_{zz,t}^{-1}\frac{\partial P_{zz,t}}{\partial\xi_1}P_{zz,t}^{-1}\Big) &\;= -P_{zz,t}^{-1}\frac{\partial P_{zz,t}}{\partial\xi_2}P_{zz,t}^{-1}\frac{\partial P_{zz,t}}{\partial\xi_1}P_{zz,t}^{-1} 
- P_{zz,t}^{-1}\frac{\partial P_{zz,t}}{\partial\xi_1}P_{zz,t}^{-1}\frac{\partial P_{zz,t}}{\partial\xi_2}P_{zz,t}^{-1}\\
&\;\qquad+  P_{zz,t}^{-1}\frac{\partial P_{zz,t}}{\partial\xi_1\partial\xi_2}P_{zz,t}^{-1}.
\end{split}
\end{equation*}
Then, we need to compute only the following second derivatives $\frac{\partial u_t}{\partial\xi_1\partial\xi_2}$ and $\frac{\partial P_{zz,t}}{\partial\xi_1\partial\xi_2}$, since all other derivatives have been already reported in Section \ref{sec:GradientKF}. By Section \ref{sec:GradientKF} we have that
\begin{equation*}
\begin{split}
\frac{\partial u_t}{\partial\xi_1\partial\xi_2} &\;= - \frac{\partial H^0}{\partial\xi_1\partial\xi_2} e -  \frac{\partial H^1}{\partial\xi_1\partial\xi_2} x_t^-  -\frac{\partial H^1}{\partial\xi_1}\frac{\partial x_t^-}{\partial\xi_2}  -\frac{\partial H^1}{\partial\xi_2}\frac{\partial x_t^-}{\partial\xi_1} - H^1\frac{\partial x_t^-}{\partial\xi_1\partial\xi_2}\\
\frac{\partial x_t^-}{\partial\xi_1\partial\xi_2} &\;= \frac{\partial\Phi^0}{\partial\xi_1\partial\xi_2} e + \frac{\partial\Phi^1}{\partial\xi_1\partial\xi_2}x_{t-1} + \frac{\partial\Phi^1}{\partial\xi_1}\frac{\partial x_{t-1}}{\partial\xi_2} + \frac{\partial\Phi^1}{\partial\xi_2}\frac{\partial x_{t-1}}{\partial\xi_1} + \Phi^1\frac{\partial x_{t-1}}{\partial\xi_1\partial\xi_2}\\
\end{split}
\end{equation*}
In both the Vasicek and the CIR case we have
\begin{equation*}
\frac{\partial\Phi^0}{\partial\xi_1\partial\xi_2} =   
\begin{pmatrix}
-\eta\Delta t^2 e^{-\kappa \Delta t} &  t e^{-\kappa \Delta t} & 0 \\
t e^{-\kappa \Delta t} & 0 & 0 \\
0 & 0 & 0 \\
\end{pmatrix}
\end{equation*}
and
\begin{equation*}
\frac{\partial\Phi^1}{\partial\xi_1\partial\xi_2} =   
\begin{pmatrix}
\Delta t^2 e^{-\kappa \Delta t} & 0 & 0 \\
0 & 0 & 0 \\
0 & 0 & 0 \\
\end{pmatrix}.
\end{equation*}
The second derivatives of $H^0$ and $H^1$ are given in Section \ref{sec:I3}. More challenging is the computation of the Hessian of $P_{zz,t}$. These involves both the factors parameters and the measurement error parameters.
\begin{equation*}
\begin{split}
\frac{\partial P_{zz,t}}{\partial\xi_1\partial\xi_2} &\;=  \frac{\partial H^1}{\partial\xi_1\partial\xi_2}P_{t}^-(H^1)' + \frac{\partial H^1}{\partial\xi_1}\frac{\partial P_{t}^-}{\partial\xi_2}(H^1)' + \frac{\partial H^1}{\partial\xi_1}P_{t}^-(\frac{\partial H^1}{\partial\xi_2})'\\
&\;\quad + \frac{\partial H^1}{\partial\xi_2}\frac{\partial P_{t}^-}{\partial\xi_1}(H^1)' + H^1\frac{\partial P_{t}^-}{\partial\xi_1\partial\xi_2}(H^1)' + H^1\frac{\partial P_{t}^-}{\partial\xi_1}(\frac{\partial H^1}{\partial\xi_2})'\\
&\;\quad  + \frac{\partial H^1}{\partial\xi_2}P_{t}^-(\frac{\partial H^1}{\partial\xi_1})' + H^1\frac{\partial P_{t}^-}{\partial\xi_2}(\frac{\partial H^1}{\partial\xi_1})'+ H^1P_{t}^-(\frac{\partial H^1}{\partial\xi_1\partial\xi_2})' + \frac{\partial\Sigma}{\partial\xi_1\partial\xi_2}
\end{split}
\end{equation*}
where, as already observed, the second derivatives of $H^1$ are given in Section \ref{sec:I3}, and the derivatives of $\Sigma$ can be computed by considering that the matrix is diagonal with all element equal to $\sigma^2_{\varepsilon}$. Then we have
\begin{equation*}
\begin{split}
\frac{\partial P_{t}^-}{\partial\xi_1\partial\xi_2} &\;=  \frac{\partial \Phi^1}{\partial\xi_1\partial\xi_2}P_{t-1}(\Phi^1)' + \frac{\partial \Phi^1}{\partial\xi_1}\frac{\partial P_{t-1}}{\partial\xi_2}(\Phi^1)' + \frac{\partial \Phi^1}{\partial\xi_1}P_{t-1}(\frac{\partial \Phi^1}{\partial\xi_2})'\\
&\;\quad + \frac{\partial \Phi^1}{\partial\xi_2}\frac{\partial P_{t-1}}{\partial\xi_1}(\Phi^1)' + \Phi^1\frac{\partial P_{t-1}}{\partial\xi_1\partial\xi_2}(\Phi^1)' + \Phi^1\frac{\partial P_{t-1}}{\partial\xi_1}(\frac{\partial \Phi^1}{\partial\xi_2})'\\
&\;\quad  + \frac{\partial \Phi^1}{\partial\xi_2}P_{t-1}(\frac{\partial \Phi^1}{\partial\xi_1})' + \Phi^1\frac{\partial P_{t-1}}{\partial\xi_2}(\frac{\partial \Phi^1}{\partial\xi_1})'+ \Phi^1P_{t-1}(\frac{\partial \Phi^1}{\partial\xi_1\partial\xi_2})' + \frac{\partial Q_t}{\partial\xi_1\partial\xi_2}
\end{split}
\end{equation*}
and . Furthermore, the filtered updates are given by
\begin{equation*}
\begin{split}
\frac{\partial x_t}{\partial\xi_1\partial\xi_2} &\;= \frac{\partial x_t^-}{\partial\xi_1\partial\xi_2} + \frac{\partial K_t}{\partial\xi_1\partial\xi_2}u_t + \frac{\partial K_t}{\partial\xi_1}\frac{\partial u_t}{\partial\xi_2} + \frac{\partial K_t}{\partial\xi_2}\frac{\partial u_t}{\partial\xi_1}+ K_t\frac{\partial u_t}{\partial\xi_1\partial\xi_2}\\
\frac{\partial P_t}{\partial\xi_1\partial\xi_2} &\;= \frac{\partial P_{t}^-}{\partial\xi_1\partial\xi_2} - \frac{\partial K_t}{\partial\xi_1\partial\xi_2}H^1P_{t}^- - \frac{\partial K_t}{\partial\xi_1}\frac{\partial H^1}{\partial\xi_2}P_{t}^- - \frac{\partial K_t}{\partial\xi_1}H^1\frac{\partial P_{t}^-}{\partial\xi_2}\\
&\;\quad - \frac{\partial K_t}{\partial\xi_2}\frac{\partial H^1}{\partial\xi_1}P_{t}^- - K_t\frac{\partial H^1}{\partial\xi_1\partial\xi_2}P_{t}^- - K_t\frac{\partial H^1}{\partial\xi_1}\frac{\partial P_{t}^-}{\partial\xi_2}\\ 
&\;\quad - \frac{\partial K_t}{\partial\xi_2}H^1\frac{\partial P_{t}^-}{\partial\xi_1} - K_t\frac{\partial H^1}{\partial\xi_2}\frac{\partial P_{t}^-}{\partial\xi_1} - K_tH^1\frac{\partial P_{t}^-}{\partial\xi_1\partial\xi_2} 
\end{split}
\end{equation*}
with
\begin{equation*}
\begin{split}
\frac{\partial K_t}{\partial\xi_1\partial\xi_2} &\;= \frac{\partial P_{t}^-}{\partial\xi_1\partial\xi_2}(H^1)'(P_{zz,t})^{-1} + \frac{\partial P_{t}^-}{\partial\xi_1}\frac{\partial (H^1)'}{\partial\xi_2}(P_{zz,t})^{-1} + \frac{\partial P_{t}^-}{\partial\xi_1}(H^1)' \frac{\partial (P_{zz,t})^{-1}}{\partial\xi_2}\\
&\;\quad + \frac{\partial P_{t}^-}{\partial\xi_2}\frac{\partial (H^1)'}{\partial\xi_1}(P_{zz,t})^{-1} + P_{t}^-\frac{\partial (H^1)'}{\partial\xi_1\partial\xi_2}(P_{zz,t})^{-1} + P_{t}^-\frac{\partial (H^1)'}{\partial\xi_1}\frac{\partial (P_{zz,t})^{-1}}{\partial\xi_2}\\
&\;\quad +\frac{\partial P_{t}^-}{\partial\xi_2}(H^1)' \frac{\partial (P_{zz,t})^{-1}}{\partial\xi_1} + P_{t}^-\frac{\partial (H^1)'}{\partial\xi_2}\frac{\partial (P_{zz,t})^{-1}}{\partial\xi_1}  - P_{t}^-(H^1)'\frac{\partial}{\partial\xi_2}\Big(P_{zz,t}^{-1}\frac{\partial P_{zz,t}}{\partial\xi_1}P_{zz,t}^{-1}\Big)
\end{split}
\end{equation*}
The second derivatives of $q_v$ for a Vasicek factor $r$ are
\begin{equation*}
\begin{split}
\frac{\partial q_v}{\partial\kappa^2} &\;= \frac{\vartheta^2}{\kappa^3} (1 - e^{-2\kappa \Delta t}) - 2\frac{\vartheta^2 \Delta t}{\kappa^2} e^{-2\kappa \Delta t} - \frac{2\vartheta^2 \Delta t^2}{\kappa} e^{-2\kappa \Delta t}\\
\frac{\partial q_v}{\partial\kappa\partial\vartheta} &\;= \frac{\partial q_v}{\partial\vartheta\partial\kappa} = -\frac{\vartheta}{\kappa^2} (1 - e^{-2\kappa \Delta t}) + \frac{2\vartheta\Delta t}{\kappa} e^{-2\kappa \Delta t}\\
\frac{\partial q_v}{\partial\vartheta^2} &\;= \frac{1}{\kappa} (1 - e^{-2\kappa \Delta t})\\
\end{split}
\end{equation*}
and all other second derivatives are zero.
The second derivatives of $Q_t$ for a CIR factor $r$ are
\begin{equation*}
\begin{split}
\frac{\partial Q_t}{\partial\xi_1\partial\xi_2} &\;= \frac{\partial q_{c1}}{\partial\xi_1\partial\xi_2} + \frac{\partial r_{t-1}}{\partial\xi_1\partial\xi_2}q_{c2} + \frac{\partial r_{t-1}}{\partial\xi_1}\frac{\partial q_{c2}}{\partial\xi_2} + \frac{\partial r_{t-1}}{\partial\xi_2}\frac{\partial q_{c2}}{\partial\xi_1} +  r_{t-1} \frac{\partial q_{c2}}{\partial\xi_1\partial\xi_2}\\
\end{split}
\end{equation*}
where
\begin{equation*}
\begin{split}
\frac{\partial q_{c1}}{\partial\kappa^2} &\;= \frac{\vartheta^2\eta}{\kappa^3} (1 - e^{-\kappa \Delta t})^2  - \frac{2\vartheta^2\eta\Delta t}{\kappa^2} (1 - e^{-\kappa \Delta t}) e^{-\kappa \Delta t} + \frac{\vartheta^2\eta\Delta t}{\kappa} (\Delta t e^{-\kappa \Delta t} - 2\Delta t e^{-2\kappa \Delta t})\\
\frac{\partial q_{c1}}{\partial\kappa\partial\eta} &\;= \frac{\partial q_{c1}}{\partial\eta\partial\kappa} =-\frac{\vartheta^2}{2\kappa^2} (1 - e^{-\kappa \Delta t})^2 + \frac{\vartheta^2\Delta t}{\kappa} (e^{-\kappa \Delta t} - e^{-2\kappa \Delta t})\\
\frac{\partial q_{c1}}{\partial\kappa\partial\vartheta} &\;= \frac{\partial q_{c1}}{\partial\vartheta\partial\kappa} = -\frac{\vartheta\eta}{\kappa^2} (1 - e^{-\kappa \Delta t})^2 + \frac{2\vartheta\eta\Delta t}{\kappa} (e^{-\kappa \Delta t} - e^{-2\kappa \Delta t})\\
\frac{\partial q_{c1}}{\partial\eta^2} &\;=0\\ 
\frac{\partial q_{c1}}{\partial\eta\partial\vartheta}&\;= \frac{\partial q_{c1}}{\partial\vartheta\partial\eta} = \frac{\vartheta}{\kappa} (1 - e^{-\kappa \Delta t})^2\\
\frac{\partial q_{c1}}{\partial\vartheta^2} &\;= \frac{\eta}{\kappa} (1 - e^{-\kappa \Delta t})^2\\
\end{split}
\end{equation*}
and
\begin{equation*}
\begin{split}
\frac{\partial q_{c2}}{\partial\kappa^2} &\;= \frac{2\vartheta^2}{\kappa^3}(e^{-\kappa \Delta t} - e^{-2\kappa \Delta t}) + \frac{2\vartheta^2}{\kappa^2}(\Delta t e^{-\kappa \Delta t} - 2 \Delta t e^{-2\kappa \Delta t}) + \frac{\vartheta^2}{\kappa}(\Delta t^2 e^{-\kappa \Delta t} - 4\Delta t^2 e^{-2\kappa \Delta t})\\
\frac{\partial q_{c2}}{\partial\kappa\partial\eta} &\;= \frac{\partial q_{c2}}{\partial\eta\partial\kappa} = \frac{\partial q_{c2}}{\partial\vartheta\partial\eta} = \frac{\partial q_{c2}}{\partial\eta\partial\xi} = 0\\
\frac{\partial q_{c2}}{\partial\kappa\partial\vartheta} &\;= \frac{\partial q_{c2}}{\partial\vartheta\partial\kappa} = -\frac{2\vartheta}{\kappa^2}(e^{-\kappa \Delta t} - e^{-2\kappa \Delta t}) - \frac{2\vartheta}{\kappa}(\Delta t e^{-\kappa \Delta t} - 2\Delta t e^{-2\kappa \Delta t})\\
\frac{\partial q_{c2}}{\partial\vartheta^2} &\;= \frac{2}{\kappa}(e^{-\kappa \Delta t} - e^{-2\kappa \Delta t}).\\
\end{split}
\end{equation*}
Furthermore, in the Vasicek case we have
\begin{equation*}
\frac{\partial P_0}{\partial\xi_1\partial\xi_2} =   
\begin{pmatrix}
\frac{\vartheta^2}{\kappa^3} &  0  & -\frac{\vartheta}{\kappa^2} \\
0 & 0 & 0 \\
-\frac{\vartheta}{\kappa^2} & 0 & \frac{1}{\kappa}\\
\end{pmatrix}
\end{equation*}
and in the CIR case we have
\begin{equation*}
\frac{\partial P_0}{\partial\xi_1\partial\xi_2} =   
\begin{pmatrix}
\frac{\vartheta^2\eta}{\kappa^3} & -\frac{\vartheta^2}{2\kappa^2} & -\frac{\vartheta\eta}{\kappa^2} \\
-\frac{\vartheta^2}{2\kappa^2} & 0 & \frac{\vartheta}{\kappa} \\
-\frac{\vartheta\eta}{\kappa^2} & \frac{\vartheta}{\kappa} & \frac{\eta}{\kappa}\\
\end{pmatrix}.
\end{equation*}
These derivatives are needed as initial point of the algorithm that compute the Hessian on the likelihood function to maximize.

\subsection{Gradient and Hessian of H0 and H1}\label{sec:I3}

As observed in equation (\ref{eq:StateSpaceModelKf_input_b}) we have that $H^0$ and $H^1$ are defined as follows
$$
H^0_{j,i} = - \frac{1}{T_j} A_i(T_j), \quad H^1_{j,i}  = - \frac{1}{T_j} B_i(T_j)
$$
therefore, to compute the gradient and the Hessian with respect to the model parameters it is enough to compute the the gradient and the Hessian of $A(t)$ and $B(t)$. 

\subsubsection{Vasicek model}

In the Vasicek case, we have
$$
z_{model} = -\frac{1}{t}\left(A(t) + B(t)r_0\right)
$$
where 
\begin{gather*}
A(t) = -\eta t - \eta g(t) - \frac{\vartheta^2}{4\kappa}\left(g(t)\right)^2 +  \frac{\vartheta^2}{2\kappa^2}\left(t + g(t)\right),\\
B(t) = g(t) = -\frac{1 - e^{-\kappa t}}{\kappa}.
\end{gather*}
It follows that the first derivatives to compute the gradient are
\begin{equation*}
\begin{split}
\frac{\partial g(t)}{\partial\kappa} &\;= \frac{1 - e^{-\kappa t} - t\kappa e^{-\kappa t} }{\kappa^2},\\
\quad \frac{\partial g(t)}{\partial\eta} &\;= \frac{\partial g(t)}{\partial\vartheta} = 0,\\
\frac{\partial A(t)}{\partial\kappa} &\;= -\eta \frac{\partial g(t)}{\partial\kappa} - \frac{\vartheta^2}{2\kappa}g(t)\frac{\partial g(t)}{\partial\kappa} + \frac{\vartheta^2}{4\kappa^2}\left(g(t)\right)^2\\ 
&\quad +  \frac{\vartheta^2}{2\kappa^2}\frac{\partial g(t)}{\partial\kappa} - \frac{\vartheta^2}{\kappa^3}\left(t + g(t)\right),\\
\frac{\partial A(t)}{\partial\eta} &\;=  - t - g(t),\\
\frac{\partial A(t)}{\partial\vartheta} &\;= - \frac{\vartheta}{2\kappa}\left(g(t)\right)^2 +  \frac{\vartheta}{\kappa^2}\left(t + g(t)\right) \\  
\end{split}
\end{equation*}
The second derivatives to compute the Hessian are
\begin{equation*}
\begin{split}
\frac{\partial g(t)}{\partial\kappa^2} &\;= \frac{t^2\kappa^2e^{-\kappa t} - 2(1 - e^{-\kappa t}) + 3t \kappa e^{-\kappa t}}{\kappa^3},\\
\end{split}
\end{equation*}
all other second derivatives of $g(t)$ are zero, and
\begin{equation*}
\begin{split}
\frac{\partial A(t)}{\partial\kappa^2} &\;= -\eta \frac{\partial g(t)}{\partial\kappa^2} + \frac{\vartheta^2}{2\kappa^2}g(t)\frac{\partial g(t)}{\partial\kappa} -  \frac{\vartheta^2}{2\kappa}\left(\frac{\partial g(t)}{\partial\kappa}\right)^2 - \frac{\vartheta^2}{2\kappa}g(t)\frac{\partial g(t)}{\partial\kappa^2} - \frac{\vartheta^2}{2\kappa^3}\left(g(t)\right)^2\\
&\:\quad + \frac{\vartheta^2}{2\kappa^2}g(t)\frac{\partial g(t)}{\partial\kappa} - \frac{\vartheta^2}{\kappa^3}\frac{\partial g(t)}{\partial\kappa} + \frac{\vartheta^2}{2\kappa^2}\frac{\partial g(t)}{\partial\kappa^2} + \frac{3\vartheta^2}{\kappa^4}\left(t + g(t)\right) - \frac{\vartheta^2}{\kappa^3}\frac{\partial g(t)}{\partial\kappa}\\
\frac{\partial A(t)}{\partial\kappa\partial\eta} &\;= \frac{\partial A(t)}{\partial\eta\partial\kappa} = -\frac{\partial g(t)}{\partial\kappa},\\
\frac{\partial A(t)}{\partial\kappa\partial\vartheta} &\;= \frac{\partial A(t)}{\partial\vartheta\partial\kappa} = - \frac{\vartheta}{\kappa}g(t)\frac{\partial g(t)}{\partial\kappa} + \frac{\vartheta}{2\kappa^2}\left(g(t)\right)^2 +  \frac{\vartheta}{\kappa^2}\frac{\partial g(t)}{\partial\kappa} - \frac{2\vartheta}{\kappa^3}\left(t + g(t)\right),\\
\frac{\partial A(t)}{\partial\eta^2} &\;= \frac{\partial A(t)}{\partial\eta\partial\vartheta} = \frac{\partial A(t)}{\partial\vartheta\partial\eta} = 0 \\
\frac{\partial A(t)}{\partial\vartheta^2} &\;= - \frac{1}{2\kappa}\left(g(t)\right)^2 +  \frac{1}{\kappa^2}\left(t + g(t)\right).\\  
\end{split}
\end{equation*}

\subsubsection{CIR model}

In the CIR case, we have
$$
z_{model} = -\frac{1}{t}\left(A(t) + B(t)r_0\right)
$$
where 
\begin{gather*}
A(t) = \frac{2\kappa\eta}{\vartheta^2} \left(\log(2) + \log(b) - \log(\kappa - b) + (\kappa + b)t/2 - \log(a e^{bt} - 1)\right),\\
B(t) = -\frac{2}{\kappa - b} \left(\frac{e^{bt} - 1}{a e^{bt} - 1}\right),\\
a = \frac{\kappa + b}{\kappa - b} \quad\text{and}\quad b = \sqrt{\kappa^2 + 2\vartheta^2}.
\end{gather*}
It follows that
$$
\frac{\partial b}{\partial\kappa} = \frac{\kappa}{\sqrt{\kappa^2 + 2\vartheta^2}}
\quad\frac{\partial b}{\partial\eta} = 0
\quad\frac{\partial b}{\partial\vartheta} = \frac{2\vartheta}{\sqrt{\kappa^2 + 2\vartheta^2}}
$$
and
\begin{equation*}
\begin{split}
\frac{\partial a}{\partial\kappa} &\;= \frac{(\kappa - b)(1+ \frac{\partial b}{\partial\kappa}) - (\kappa + b)(1 - \frac{\partial b}{\partial\kappa})}{(\kappa - b)^2} = \frac{2(\kappa\frac{\partial b}{\partial\kappa} - b)}{(\kappa - b)^2}\\
\quad\frac{\partial a}{\partial\eta} &\;= 0\\
\quad\frac{\partial a}{\partial\vartheta} &\;= \frac{(\kappa - b)\frac{\partial b}{\partial\vartheta} + (\kappa + b)\frac{\partial b}{\partial\vartheta}}{(\kappa - b)^2} = \frac{2\kappa\frac{\partial b}{\partial\vartheta}}{(\kappa - b)^2}\\
\end{split}
\end{equation*}

Thus, by setting 
\begin{equation*}
\begin{split}
f &\;= \log(2) + \log(b) - \log(\kappa - b) + (\kappa + b)t/2 - w\\
w &\;= \log(a e^{bt} - 1),\\
\end{split}
\end{equation*}
the equalities
\begin{equation*}
\begin{split}
\frac{\partial f}{\partial\kappa}&\;= \frac{1}{b}\frac{\partial b}{\partial\kappa} - \frac{1}{k - b}\left(1 - \frac{\partial b}{\partial\kappa}\right) + \left(1 +\frac{\partial b}{\partial\kappa}\right)\frac{t}{2} - \frac{\partial w}{\partial\kappa}\\
\frac{\partial f}{\partial\eta} &\;= 0\\
\frac{\partial f}{\partial\vartheta} &\;= \frac{1}{b}\frac{\partial b}{\partial\vartheta} + \frac{1}{k - b}\frac{\partial b}{\partial\vartheta} + \frac{t}{2}\frac{\partial b}{\partial\vartheta} - \frac{\partial w}{\partial\vartheta}\\
\frac{\partial w}{\partial\xi}&\;= \frac{\frac{\partial a}{\partial\xi}e^{bt} + at e^{bt}\frac{\partial b}{\partial\xi}}{a e^{bt} - 1}
\end{split}
\end{equation*}
follow and we obtain for $A(t)$ the following partial derivatives
\begin{equation*}
\begin{split}
\frac{\partial A(t)}{\partial\kappa}&\;= \frac{2\eta}{\vartheta^2}f + \frac{2\kappa\eta}{\vartheta^2} \frac{\partial f}{\partial\kappa}\\
\frac{\partial A(t)}{\partial\eta}&\;= \frac{2\kappa}{\vartheta^2} f\\
\frac{\partial A(t)}{\partial\vartheta}&\;= -\frac{4\kappa\eta}{\vartheta^3}f + \frac{2\kappa\eta}{\vartheta^2} \frac{\partial f}{\partial\vartheta},
\end{split}
\end{equation*}
and for $B(t)$
\begin{equation*}
\begin{split}
\frac{\partial B(t)}{\partial\kappa}&\;= \frac{2h}{(\kappa - b)^2}\left(1 -\frac{\partial b}{\partial\kappa}\right) - \frac{2}{\kappa - b}\frac{\partial h}{\partial\kappa}\\
\frac{\partial B(t)}{\partial\eta}&\;= 0\\
\frac{\partial B(t)}{\partial\vartheta}&\;= -\frac{2h}{(\kappa - b)^2}\frac{\partial b}{\partial\vartheta} - \frac{2}{\kappa - b}\frac{\partial h}{\partial\vartheta}\\
\end{split}
\end{equation*}
where
$$
h = \frac{e^{bt} - 1}{a e^{bt} - 1}
$$
and for the generic parameter $\xi \in\{\kappa,\vartheta\}$
$$
\frac{\partial h}{\partial\xi } = \frac{e^{bt}\left(-t\frac{\partial b}{\partial\xi} - e^{bt}\frac{\partial a}{\partial\xi} + \frac{\partial a}{\partial\xi} + at\frac{\partial b}{\partial\xi}\right)}{(a e^{bt} - 1)^2}.
$$
A little bit more tedious are the computation of the second derivatives. More in details, we have
\begin{equation*}
\begin{split}
\frac{\partial b}{\partial\kappa^2}&\; = \frac{2\vartheta^2}{(\kappa^2 + 2\vartheta^2)^{\frac{3}{2}}}\\
\frac{\partial b}{\partial\kappa\partial\vartheta}&\; = \frac{\partial b}{\partial\vartheta\partial\kappa} = -\frac{2\kappa\vartheta}{(\kappa^2 + 2\vartheta^2)^{\frac{3}{2}}}\\
\frac{\partial b}{\partial\vartheta^2}&\; =  \frac{2\kappa^2}{(\kappa^2 + 2\vartheta^2)^{\frac{3}{2}}}\\
\end{split}
\end{equation*}
with all other second derivatives equal to zero, and
\begin{equation*}
\begin{split}
\frac{\partial a}{\partial\kappa^2} &\;= \frac{2k\frac{\partial b}{\partial\kappa^2}(\kappa - b) - 4(1 - \frac{\partial b}{\partial\kappa})(k\frac{\partial b}{\partial\kappa} - b)}{(\kappa - b)^3}\\
\frac{\partial a}{\partial\kappa\partial\vartheta} &\;= \frac{\partial a}{\partial\vartheta\partial\kappa} = \frac{2(\kappa\frac{\partial b}{\partial\kappa\partial\vartheta})(\kappa - b) + 2\frac{\partial b}{\partial\vartheta}(2\kappa\frac{\partial b}{\partial\kappa} - b - k)}{(\kappa - b)^3}\\
\frac{\partial a}{\partial\vartheta^2}&\;= \frac{2\kappa\frac{\partial b}{\partial\vartheta^2}(k - b) + 4\kappa(\frac{\partial b}{\partial\vartheta})^2}{(\kappa - b)^3}\\
\end{split}
\end{equation*}
with all other second derivatives equal to zero. Then, by setting 
\begin{equation*}
\begin{split}
\frac{\partial f}{\partial\kappa^2}&\;= -\frac{1}{b^2}\left(\frac{\partial b}{\partial\kappa}\right)^2 + \frac{1}{b}\frac{\partial b}{\partial\kappa^2} + \frac{1}{(k - b)^2}\left(1 - \frac{\partial b}{\partial\kappa}\right)^2 + \frac{1}{k - b}\frac{\partial b}{\partial\kappa^2} + \frac{t}{2}\frac{\partial b}{\partial\kappa^2} - \frac{\partial w}{\partial\kappa^2}\\
\frac{\partial f}{\partial\kappa\partial\vartheta}&\;= \frac{\partial f}{\partial\vartheta\partial\kappa} = -\frac{1}{b^2}\frac{\partial b}{\partial\kappa}\frac{\partial b}{\partial\vartheta} + \frac{1}{b}\frac{\partial b}{\partial\kappa\partial\vartheta} - \frac{1}{(k - b)^2}\left(1 - \frac{\partial b}{\partial\kappa}\right)\frac{\partial b}{\partial\vartheta}\\ &\;\quad +\frac{1}{k - b}\frac{\partial b}{\partial\kappa\partial\vartheta} + \frac{t}{2}\frac{\partial b}{\partial\kappa\partial\vartheta} - \frac{\partial w}{\partial\kappa\partial\vartheta}\\
\frac{\partial f}{\partial\vartheta^2}&\;= -\frac{1}{b^2}\left(\frac{\partial b}{\partial\vartheta}\right)^2 + \frac{1}{b}\frac{\partial b}{\partial\vartheta^2} + \frac{1}{(k - b)^2}\left(\frac{\partial b}{\partial\vartheta}\right)^2 + \frac{1}{k - b}\frac{\partial b}{\partial\vartheta^2} + \frac{t}{2}\frac{\partial b}{\partial\vartheta^2} - \frac{\partial w}{\partial\vartheta^2}
\end{split}
\end{equation*}
with
\begin{equation*}
\begin{split}
\frac{\partial w}{\partial\kappa^2} &\;= \frac{\frac{\partial a}{\partial\kappa^2}e^{bt} + 2\frac{\partial a}{\partial\kappa}t e^{bt}\frac{\partial b}{\partial\kappa} + a t^2 e^{bt}\left(\frac{\partial b}{\partial\kappa}\right)^2 + ate^{bt}\frac{\partial b}{\partial\kappa^2} }{(a e^{bt} - 1)}\\
&\;\quad - \frac{\left(\frac{\partial a}{\partial\kappa}e^{bt} + at e^{bt}\frac{\partial b}{\partial\kappa}\right)^2}{(a e^{bt} - 1)^2}\\
\frac{\partial w}{\partial\kappa\partial\vartheta} &\;=  \frac{\partial w}{\partial\vartheta\partial\kappa} = \frac{\frac{\partial a}{\partial\kappa\partial\vartheta}e^{bt} + \frac{\partial a}{\partial\kappa}t e^{bt}\frac{\partial b}{\partial\vartheta} + \frac{\partial a}{\partial\vartheta} t e^{bt}\frac{\partial b}{\partial\kappa} + a t^2 e^{bt}\frac{\partial b}{\partial\kappa}\frac{\partial b}{\partial\vartheta} + ate^{bt}\frac{\partial b}{\partial\kappa}\frac{\partial b}{\partial\vartheta} }{(a e^{bt} - 1)}\\
&\;\quad - \frac{\left(\frac{\partial a}{\partial\kappa}e^{bt} + at e^{bt}\frac{\partial b}{\partial\kappa}\right) \left(\frac{\partial a}{\partial\vartheta}e^{bt} + at e^{bt}\frac{\partial b}{\partial\vartheta}\right)}{(a e^{bt} - 1)^2}\\
\frac{\partial w}{\partial\vartheta^2}&\;= \frac{\frac{\partial a}{\partial\vartheta^2}e^{bt} + 2\frac{\partial a}{\partial\vartheta}t e^{bt}\frac{\partial b}{\partial\vartheta} + a t^2 e^{bt}\left(\frac{\partial b}{\partial\vartheta}\right)^2 + ate^{bt}\frac{\partial b}{\partial\vartheta^2} }{(a e^{bt} - 1)}\\
&\;\quad - \frac{\left(\frac{\partial a}{\partial\vartheta}e^{bt} + at e^{bt}\frac{\partial b}{\partial\vartheta}\right)^2}{(a e^{bt} - 1)^2}\\
\end{split}
\end{equation*}
we obtain for $A(t)$ the following partial derivatives
\begin{equation*}
\begin{split}
\frac{\partial A(t)}{\partial\kappa^2}&\;= \frac{4\eta}{\vartheta^2}\frac{\partial f}{\partial\kappa} + \frac{2\kappa\eta}{\vartheta^2} \frac{\partial f}{\partial\kappa^2}\\
\frac{\partial A(t)}{\partial\kappa\partial\vartheta}&\;= \frac{\partial A(t)}{\partial\vartheta\partial\kappa} = -\frac{4\eta}{\vartheta^3} f + \frac{2\eta}{\vartheta^2}\frac{\partial f}{\partial\vartheta} - \frac{4\kappa\eta}{\vartheta^3}\frac{\partial f}{\partial\kappa} + \frac{2\kappa\eta}{\vartheta^2} \frac{\partial f}{\partial\kappa\partial\vartheta}\\
\frac{\partial A(t)}{\partial\vartheta^2}&\;= \frac{12\kappa\eta}{\vartheta^4}f - \frac{8 \kappa\eta}{\vartheta^3}\frac{\partial f}{\partial\vartheta} + \frac{2\kappa\eta}{\vartheta^2} \frac{\partial f}{\partial\vartheta^2}
\end{split}
\end{equation*}
and for $B(t)$
\begin{equation*}
\begin{split}
\frac{\partial B(t)}{\partial\kappa^2}&\;=  \frac{\partial h}{\partial\kappa}\frac{2}{(\kappa - b)^2}\left(1 -\frac{\partial b}{\partial\kappa}\right) - \frac{4h}{(\kappa - b)^3}\left(1- \frac{\partial b}{\partial\kappa}\right)^2 - \frac{2h}{(\kappa - b)^2}\frac{\partial b}{\partial\kappa^2}\\ 
&\;\quad + \frac{2}{(\kappa - b)^2}\frac{\partial h}{\partial\kappa}\left(1 -\frac{\partial b}{\partial\kappa}\right) - \frac{2}{\kappa - b}\frac{\partial h}{\partial\kappa^2}\\
\frac{\partial B(t)}{\partial\kappa\partial\vartheta}&\;= \frac{\partial B(t)}{\partial\vartheta\partial\kappa} = \frac{\partial h}{\partial\vartheta}\frac{2}{(\kappa - b)^2}\left(1 -\frac{\partial b}{\partial\kappa}\right) + \frac{4h}{(\kappa - b)^3}\left(1 - \frac{\partial b}{\partial\kappa}\right)\frac{\partial b}{\partial\vartheta} - \frac{2h}{(\kappa - b)^2}\frac{\partial b}{\partial\kappa\partial\vartheta}\\ 
&\;\quad - \frac{2}{(\kappa - b)^2}\frac{\partial h}{\partial\kappa}\frac{\partial b}{\partial\vartheta} - \frac{2}{\kappa - b}\frac{\partial h}{\partial\kappa\partial\vartheta}\\
\frac{\partial B(t)}{\partial\vartheta^2}&\;= -\frac{\partial h}{\partial\vartheta}\frac{2}{(\kappa - b)^2}\frac{\partial b}{\partial\vartheta} - \frac{4h}{(\kappa - b)^3}\left(\frac{\partial b}{\partial\vartheta}\right)^2 - \frac{2h}{(\kappa - b)^2}\frac{\partial b}{\partial\vartheta^2}\\ 
&\;\quad - \frac{2}{(\kappa - b)^2}\frac{\partial h}{\partial\vartheta}\frac{\partial b}{\partial\vartheta} - \frac{2}{\kappa - b}\frac{\partial h}{\partial\vartheta^2}\\
\end{split}
\end{equation*}
with
\begin{equation*}
\begin{split}
\frac{\partial h}{\partial\kappa^2} &\;= \frac{t\frac{\partial b}{\partial\kappa}e^{bt}\left(-t\frac{\partial b}{\partial\kappa} - e^{bt}\frac{\partial a}{\partial\kappa} + \frac{\partial a}{\partial\kappa} + at\frac{\partial b}{\partial\kappa}\right)}{(a e^{bt} - 1)^2}\\ 
&\;\quad +  \frac{e^{bt}\left(-t\frac{\partial b}{\partial\kappa^2} - t\frac{\partial b}{\partial\kappa}e^{bt}\frac{\partial a}{\partial\kappa} - e^{bt}\frac{\partial a}{\partial\kappa^2}  + \frac{\partial a}{\partial\kappa^2} + \frac{\partial a}{\partial\kappa}t\frac{\partial b}{\partial\kappa} + a t \frac{\partial b}{\partial\kappa^2}\right)}{(a e^{bt} - 1)^2}\\
&\;\quad - \frac{2e^{bt}\left(-t\frac{\partial b}{\partial\kappa} - e^{bt}\frac{\partial a}{\partial\kappa} + \frac{\partial a}{\partial\kappa} + at\frac{\partial b}{\partial\kappa}\right)\left(\frac{\partial a}{\partial\kappa}e^{bt} + ate^{bt}\frac{\partial b}{\partial\kappa} \right) }{(a e^{bt} - 1)^3}\\
\frac{\partial h}{\partial\kappa\partial\vartheta} &\;=  \frac{t\frac{\partial b}{\partial\vartheta}e^{bt}\left(-t\frac{\partial b}{\partial\kappa} - e^{bt}\frac{\partial a}{\partial\kappa} + \frac{\partial a}{\partial\kappa} + at\frac{\partial b}{\partial\kappa}\right)}{(a e^{bt} - 1)^2}\\  
&\;\quad +  \frac{e^{bt}\left(-t\frac{\partial b}{\partial\kappa\partial\vartheta} - t\frac{\partial b}{\partial\vartheta}e^{bt}\frac{\partial a}{\partial\kappa} - e^{bt}\frac{\partial a}{\partial\kappa\partial\vartheta} + \frac{\partial a}{\partial\kappa\partial\vartheta} + \frac{\partial a}{\partial\vartheta}t\frac{\partial b}{\partial\kappa} + a t \frac{\partial b}{\partial\kappa\partial\vartheta}\right)}{(a e^{bt} - 1)^2}\\
&\;\quad - \frac{2e^{bt}\left(-t\frac{\partial b}{\partial\kappa} - e^{bt}\frac{\partial a}{\partial\kappa} + \frac{\partial a}{\partial\kappa} + at\frac{\partial b}{\partial\kappa}\right)\left(\frac{\partial a}{\partial\vartheta}e^{bt} + ate^{bt}\frac{\partial b}{\partial\vartheta} \right) }{(a e^{bt} - 1)^3}\\
\frac{\partial h}{\partial\vartheta^2}&\;= \frac{t\frac{\partial b}{\partial\vartheta}e^{bt}\left(-t\frac{\partial b}{\partial\vartheta} - e^{bt}\frac{\partial a}{\partial\vartheta} + \frac{\partial a}{\partial\vartheta} + at\frac{\partial b}{\partial\vartheta}\right)}{(a e^{bt} - 1)^2}\\  
&\;\quad +  \frac{e^{bt}\left(-t\frac{\partial b}{\partial\vartheta^2} - t\frac{\partial b}{\partial\vartheta}e^{bt}\frac{\partial a}{\partial\vartheta} - e^{bt}\frac{\partial a}{\partial\vartheta^2} + \frac{\partial a}{\partial\vartheta^2} + \frac{\partial a}{\partial\vartheta}t\frac{\partial b}{\partial\vartheta} + a t \frac{\partial b}{\partial\vartheta^2}\right)}{(a e^{bt} - 1)^2}\\
&\;\quad - \frac{2e^{bt}\left(-t\frac{\partial b}{\partial\vartheta} - e^{bt}\frac{\partial a}{\partial\vartheta} + \frac{\partial a}{\partial\vartheta} + at\frac{\partial b}{\partial\vartheta}\right)\left(\frac{\partial a}{\partial\vartheta}e^{bt} + ate^{bt}\frac{\partial b}{\partial\vartheta} \right) }{(a e^{bt} - 1)^3}\\
\end{split}
\end{equation*}

\newpage


\newpage



\begin{sidewaystable}
\begin{center}
\begin{footnotesize}
\begin{tabular}{@{}lcccccccccccccccc|c@{}}
\toprule
& \multicolumn{15}{c}{Years to maturity}\\ 
\midrule																													
& &	0.25	&	0.50	&	1	&	2	&	3	&	4	&	5	&	6	&	7	&	8	&	9	&	10	&	15	&	20	&	30 & total	\\
\midrule																								
\multirow{2}{*}{1-CIR}	&	APE	&	14.55	&	12.68	&	9.34	&	5.20	&	3.72	&	4.86	&	6.56	&	7.40	&	8.37	&	9.37	&	10.61	&	11.19	&	14.92	&	15.49	&	13.00	&	9.88	\\
	&	RMSE	&	72.23	&	59.06	&	42.84	&	24.64	&	20.72	&	24.00	&	30.74	&	34.74	&	39.80	&	42.69	&	46.30	&	50.31	&	69.14	&	70.80	&	61.41	&	48.96	\\
\midrule																																			
\multirow{2}{*}{1-Vasicek}	&	APE	&	14.50	&	12.71	&	9.31	&	4.91	&	3.60	&	4.81	&	6.52	&	7.37	&	8.33	&	9.27	&	10.50	&	11.10	&	14.86	&	15.43	&	13.01	&	9.85	\\
	&	RMSE	&	71.96	&	59.02	&	42.55	&	24.10	&	20.37	&	23.84	&	30.80	&	34.94	&	40.03	&	42.83	&	46.38	&	50.47	&	69.39	&	70.94	&	61.58	&	48.98	\\
\midrule																																			
\multirow{2}{*}{2-CIR}	&	APE	&	7.85	&	6.42	&	5.91	&	5.50	&	5.10	&	4.62	&	4.77	&	4.73	&	4.55	&	4.07	&	5.01	&	5.22	&	8.64	&	8.76	&	9.36	&	5.70	\\
	&	RMSE	&	44.82	&	36.94	&	35.63	&	33.54	&	31.84	&	29.60	&	29.76	&	28.84	&	27.98	&	26.74	&	28.49	&	29.20	&	39.66	&	41.17	&	45.45	&	34.51	\\
\midrule																																			
\multirow{2}{*}{2-Vasicek}	&	APE	&	4.71	&	3.29	&	2.75	&	3.42	&	3.35	&	3.05	&	2.90	&	2.87	&	2.91	&	2.20	&	2.35	&	2.24	&	3.44	&	4.72	&	5.04	&	3.25	\\
	&	RMSE	&	29.75	&	22.15	&	22.16	&	23.72	&	22.37	&	19.70	&	18.78	&	17.81	&	18.17	&	16.21	&	16.55	&	16.00	&	19.00	&	25.20	&	23.39	&	21.06	\\
\midrule																																			
\multirow{2}{*}{1-CIR + 1-Vasicek}	&	APE	&	4.74	&	3.29	&	2.77	&	3.41	&	3.33	&	3.07	&	2.87	&	2.82	&	2.92	&	2.26	&	2.40	&	2.31	&	3.41	&	4.79	&	5.21	&	3.28	\\
	&	RMSE	&	30.02	&	22.29	&	22.28	&	23.85	&	22.45	&	19.81	&	18.74	&	17.71	&	18.19	&	16.40	&	16.76	&	16.23	&	18.81	&	25.41	&	23.53	&	21.17	\\
\midrule																																			
\multirow{2}{*}{3-CIR}	&	APE	&	5.82	&	3.96	&	5.10	&	5.60	&	5.69	&	4.97	&	5.28	&	5.66	&	5.65	&	4.56	&	4.11	&	4.54	&	7.57	&	8.86	&	5.86	&	5.15	\\
	&	RMSE	&	32.11	&	20.80	&	24.61	&	25.87	&	25.74	&	23.30	&	24.55	&	25.08	&	24.63	&	21.63	&	20.35	&	21.64	&	32.08	&	38.49	&	28.21	&	26.38	\\
\midrule																																			
\multirow{2}{*}{3-Vasicek}	&	APE	&	2.25	&	1.80	&	3.09	&	2.49	&	2.24	&	2.07	&	1.96	&	2.27	&	2.71	&	2.25	&	2.28	&	2.13	&	3.03	&	3.99	&	4.27	&	2.61	\\
	&	RMSE	&	20.60	&	19.15	&	22.39	&	20.23	&	19.01	&	17.41	&	16.46	&	16.48	&	17.67	&	16.21	&	16.44	&	15.47	&	18.16	&	22.65	&	21.53	&	18.79	\\
\midrule																																			
\multirow{2}{*}{2-CIR + 1-Vasicek}	&	APE	&	2.27	&	1.90	&	2.79	&	2.47	&	2.27	&	2.18	&	2.17	&	2.52	&	2.68	&	2.11	&	2.57	&	2.33	&	4.30	&	4.45	&	7.14	&	2.92	\\
	&	RMSE	&	25.89	&	24.53	&	26.57	&	25.19	&	24.08	&	22.87	&	22.24	&	22.19	&	22.42	&	20.76	&	21.39	&	20.48	&	25.75	&	27.10	&	32.81	&	24.48	\\

\bottomrule
\end{tabular}
\caption[{\it Italian term structure calibration error}]{\label{tab:CalibrationError}\footnotesize Calibration error of the multi-factor Gaussian models. We report, for each model and for each maturity, the bond term structure calibration error in the period from May 31, 2003 to May 31, 2014. The average percentage error (APE) in percentage points and the root mean square error (RMSE) in basis points are reported.}
\end{footnotesize}
\end{center}
\end{sidewaystable}
\begin{sidewaystable}
\begin{center}
\begin{footnotesize}
\begin{tabular}{@{}lcccccccccclc @{}}
\toprule
																									
	&	$\kappa_1$	&	$\eta_1$	&	$\vartheta_1$	&	$\kappa_2$	&	$\eta_2$	&	$\vartheta_2$	&	$\kappa_3$	&	$\eta_3$	&	$\vartheta_3$	&	$\sigma_{\varepsilon}$	&	&	\\
																																												
\midrule																																															
\multirow{3}{*}{1-CIR}	&		0.2110		&		0.0657		&		0.0995		&				&				&				&				&				&				&		0.0765		&	LL	&		1.679e+5		\\
	&	{\scriptsize {\it	-8.15e-2	}}	&	{\scriptsize {\it	-5.68e-1	}}	&	{\scriptsize {\it	1.64e-2	}}	&	{\scriptsize {\it		}}	&	{\scriptsize {\it		}}	&	{\scriptsize {\it		}}	&	{\scriptsize {\it		}}	&	{\scriptsize {\it		}}	&	{\scriptsize {\it		}}	&	{\scriptsize {\it	9.41e-2	}}	&	opt.con.	&	{\scriptsize {\it	5.68e-1	}}	\\
	&	{\scriptsize {\it	-7.50e-8	}}	&	{\scriptsize {\it	3.07e-9	}}	&	{\scriptsize {\it	-4.46e-10	}}	&	{\scriptsize {\it		}}	&	{\scriptsize {\it		}}	&	{\scriptsize {\it		}}	&	{\scriptsize {\it		}}	&	{\scriptsize {\it		}}	&	{\scriptsize {\it		}}	&	{\scriptsize {\it	6.88e-8	}}	&	AIC	&	{\scriptsize {\it	-3.36e+5	}}	\\
\midrule																																															
\multirow{3}{*}{1-Vasicek}	&		0.2433		&		0.0611		&		0.0124		&				&				&				&				&				&				&		0.0766		&	LL	&		1.678e+5		\\
	&	{\scriptsize {\it	9.65e-7	}}	&	{\scriptsize {\it	5.77e-6	}}	&	{\scriptsize {\it	-3.50e-7	}}	&	{\scriptsize {\it		}}	&	{\scriptsize {\it		}}	&	{\scriptsize {\it		}}	&	{\scriptsize {\it		}}	&	{\scriptsize {\it		}}	&	{\scriptsize {\it		}}	&	{\scriptsize {\it	-3.74e-7	}}	&	opt.con.	&	{\scriptsize {\it	5.77e-6	}}	\\
	&	{\scriptsize {\it	1.45e-6	}}	&	{\scriptsize {\it	8.93e-9	}}	&	{\scriptsize {\it	1.14e-8	}}	&	{\scriptsize {\it		}}	&	{\scriptsize {\it		}}	&	{\scriptsize {\it		}}	&	{\scriptsize {\it		}}	&	{\scriptsize {\it		}}	&	{\scriptsize {\it		}}	&	{\scriptsize {\it	6.90e-8	}}	&	AIC	&	{\scriptsize {\it	-3.36e+5	}}	\\
\midrule																																															
\multirow{3}{*}{2-CIR}	&		0.3827		&		0.0425		&		0.0928		&		0.0092		&		0.1000		&		0.0366		&				&				&				&		0.0377		&	LL	&		1.906e+5		\\
	&	{\scriptsize {\it	-1.46e+4	}}	&	{\scriptsize {\it	-1.35e+6	}}	&	{\scriptsize {\it	2.74e+4	}}	&	{\scriptsize {\it	2.44e+6	}}	&	{\scriptsize {\it	2.48e+5	}}	&	{\scriptsize {\it	-1.89e+5	}}	&	{\scriptsize {\it		}}	&	{\scriptsize {\it		}}	&	{\scriptsize {\it		}}	&	{\scriptsize {\it	-6.91e+4	}}	&	opt.con.	&	{\scriptsize {\it	2.44e+6	}}	\\
	&	{\scriptsize {\it	-7.46e-9	}}	&	{\scriptsize {\it	-1.84e-11	}}	&	{\scriptsize {\it	-2.51e-9	}}	&	{\scriptsize {\it	-6.80e-12	}}	&	{\scriptsize {\it	-9.99e-9	}}	&	{\scriptsize {\it	-5.57e-11	}}	&	{\scriptsize {\it		}}	&	{\scriptsize {\it		}}	&	{\scriptsize {\it		}}	&	{\scriptsize {\it	1.44e-8	}}	&	AIC	&	{\scriptsize {\it	-3.81e+5	}}	\\
\midrule																																															
\multirow{3}{*}{2-Vasicek}	&		0.2666		&		0.1000		&		0.0113		&		0.0535		&				&		0.0175		&				&				&				&		0.0257		&	LL	&		2.125e+5		\\
	&	{\scriptsize {\it	1.82e-3	}}	&	{\scriptsize {\it	-2.23e+4	}}	&	{\scriptsize {\it	-5.88e-3	}}	&	{\scriptsize {\it	1.77e-2	}}	&	{\scriptsize {\it		}}	&	{\scriptsize {\it	-5.91e-2	}}	&	{\scriptsize {\it		}}	&	{\scriptsize {\it		}}	&	{\scriptsize {\it		}}	&	{\scriptsize {\it	-2.42e-2	}}	&	opt.con.	&	{\scriptsize {\it	5.91e-2	}}	\\
	&	{\scriptsize {\it	1.18e-6	}}	&	{\scriptsize {\it	1.74e-8	}}	&	{\scriptsize {\it	3.57e-9	}}	&	{\scriptsize {\it	1.64e-9	}}	&	{\scriptsize {\it		}}	&	{\scriptsize {\it	1.16e-10	}}	&	{\scriptsize {\it		}}	&	{\scriptsize {\it		}}	&	{\scriptsize {\it		}}	&	{\scriptsize {\it	8.17e-9	}}	&	AIC	&	{\scriptsize {\it	-4.25e+5	}}	\\
\midrule																																															
\multirow{3}{*}{1-CIR + 1-Vasicek}	&		0.0319		&		0.1000		&		0.0574		&		0.2338		&				&		0.0110		&				&				&				&		0.0259		&	LL	&		2.125e+5		\\
	&	{\scriptsize {\it	2.14e+0	}}	&	{\scriptsize {\it	-6.43e+3	}}	&	{\scriptsize {\it	-1.38e+0	}}	&	{\scriptsize {\it	-2.56e-1	}}	&	{\scriptsize {\it		}}	&	{\scriptsize {\it	-1.89e+0	}}	&	{\scriptsize {\it		}}	&	{\scriptsize {\it		}}	&	{\scriptsize {\it		}}	&	{\scriptsize {\it	1.56e-1	}}	&	opt.con.	&	{\scriptsize {\it	2.14e+0	}}	\\
	&	{\scriptsize {\it	1.86e-11	}}	&	{\scriptsize {\it	2.34e-8	}}	&	{\scriptsize {\it	4.62e-11	}}	&	{\scriptsize {\it	1.73e-7	}}	&	{\scriptsize {\it		}}	&	{\scriptsize {\it	1.41e-9	}}	&	{\scriptsize {\it		}}	&	{\scriptsize {\it		}}	&	{\scriptsize {\it		}}	&	{\scriptsize {\it	8.26e-9	}}	&	AIC	&	{\scriptsize {\it	-4.25e+5	}}	\\
\midrule																																															
\multirow{3}{*}{3-CIR}	&		0.0826		&		0.1000		&		0.2521		&		0.0001		&		0.0001		&		0.0428		&		0.6565		&		0.0160		&		0.1390		&		0.0340		&	LL	&		1.937e+5		\\
	&	{\scriptsize {\it	-4.10e+5	}}	&	{\scriptsize {\it	-4.96e+5	}}	&	{\scriptsize {\it	2.36e+5	}}	&	{\scriptsize {\it	-5.25e+4	}}	&	{\scriptsize {\it	4.37e+3	}}	&	{\scriptsize {\it	-7.39e+4	}}	&	{\scriptsize {\it	-7.82e+2	}}	&	{\scriptsize {\it	-1.33e+5	}}	&	{\scriptsize {\it	1.22e+3	}}	&	{\scriptsize {\it	2.16e+5	}}	&	opt.con.	&	{\scriptsize {\it	4.96e+5	}}	\\
	&	{\scriptsize {\it	-2.57e-13	}}	&	{\scriptsize {\it	-4.37e-13	}}	&	{\scriptsize {\it	-4.54e-13	}}	&	{\scriptsize {\it	-6.06e-15	}}	&	{\scriptsize {\it	-3.67e-8	}}	&	{\scriptsize {\it	-9.00e-10	}}	&	{\scriptsize {\it	-1.31e-8	}}	&	{\scriptsize {\it	-7.71e-9	}}	&	{\scriptsize {\it	3.04e-9	}}	&	{\scriptsize {\it	1.96e-8	}}	&	AIC	&	{\scriptsize {\it	-3.87e+5	}}	\\
\midrule																																															
\multirow{3}{*}{3-Vasicek}	&		0.2267		&		0.1000		&		0.0108		&		0.0373		&		0.0141		&		2.0562		&				&		0.0131		&				&		0.0203		&	LL	&		2.216e+5		\\
	&	{\scriptsize {\it	1.49e-2	}}	&	{\scriptsize {\it	-2.58e+4	}}	&	{\scriptsize {\it	4.86e-1	}}	&	{\scriptsize {\it	-1.72e-1	}}	&	{\scriptsize {\it	1.03e+0	}}	&	{\scriptsize {\it	-1.16e-3	}}	&	{\scriptsize {\it		}}	&	{\scriptsize {\it	-5.01e-1	}}	&	{\scriptsize {\it		}}	&	{\scriptsize {\it	3.42e-1	}}	&	opt.con.	&	{\scriptsize {\it	1.23e+0	}}	\\
	&	{\scriptsize {\it	9.07e-7	}}	&	{\scriptsize {\it	3.51e-8	}}	&	{\scriptsize {\it	4.92e-9	}}	&	{\scriptsize {\it	5.96e-10	}}	&	{\scriptsize {\it	6.62e-11	}}	&	{\scriptsize {\it	1.66e-3	}}	&	{\scriptsize {\it		}}	&	{\scriptsize {\it	1.80e-8	}}	&	{\scriptsize {\it		}}	&	{\scriptsize {\it	5.26e-9	}}	&	AIC	&	{\scriptsize {\it	-4.43e+5	}}	\\
\midrule																																															
\multirow{3}{*}{2-CIR + 1-Vasicek}	&		1.4988		&		0.0348		&		0.0956		&		0.0089		&		0.1000		&		0.0396		&		0.3260		&				&		0.0136		&		0.0015		&	LL	&		2.165e+5		\\
	&	{\scriptsize {\it	-3.94e+2	}}	&	{\scriptsize {\it	3.74e+5	}}	&	{\scriptsize {\it	3.40e+3	}}	&	{\scriptsize {\it	1.22e+6	}}	&	{\scriptsize {\it	1.32e+5	}}	&	{\scriptsize {\it	-2.75e+5	}}	&	{\scriptsize {\it	-1.47e+3	}}	&	{\scriptsize {\it		}}	&	{\scriptsize {\it	-4.62e+4	}}	&	{\scriptsize {\it	3.77e+3	}}	&	opt.con.	&	{\scriptsize {\it	1.22e+6	}}	\\
	&	{\scriptsize {\it	-1.66e-9	}}	&	{\scriptsize {\it	-1.66e-9	}}	&	{\scriptsize {\it	1.67e-12	}}	&	{\scriptsize {\it	1.73e-12	}}	&	{\scriptsize {\it	1.76e-8	}}	&	{\scriptsize {\it	2.72e-11	}}	&	{\scriptsize {\it	1.56e-6	}}	&	{\scriptsize {\it		}}	&	{\scriptsize {\it	2.70e-9	}}	&	{\scriptsize {\it	2.33e-7	}}	&	AIC	&	{\scriptsize {\it	-4.33e+5	}}	\\

\bottomrule
\end{tabular}
\caption[{\it Italian term structure estimated parameters}]{\label{tab:EstimatedParameters}\footnotesize Estimated parameters of the multi-factor Gaussian models in the period from May 31, 2003 to May 31, 2014. We report, for each model the estimated parameters, the corresponding value of the gradient and the square of the standard error. The value of the likelihood (LL), of the first order optimality condition at the optimal point (opt.con), and of the AIC are shown in the last column.}
\end{footnotesize}
\end{center}
\end{sidewaystable}

\processdelayedfloats

%

\processdelayedfloats

\processdelayedfloats

\end{document}